\documentclass[aps, pra,twocolumn, showpacs,superscriptaddress,10pt,floatfix]{revtex4-2}

\usepackage{graphicx}% Include figure files
\usepackage{dcolumn}% Align table columns on decimal point
\usepackage{bm}% bold math
\usepackage{hyperref}% add hypertext capabilities
\usepackage{xcolor}
\usepackage{mathrsfs}
\usepackage{physics}
\usepackage[page]{appendix}
\usepackage{tikz}
\usepackage{bbold}
\usepackage{amsmath}
\usepackage{qcircuit}
\usepackage{soul}
\usepackage{listings}

\newcommand{\code}[1]{{[[#1]]}}

\begin{document}

\title{Simultaneous Discovery of Quantum Error Correction Codes and Encoders with a Noise-Aware Reinforcement Learning Agent}
\author{Jan Olle}
\email{jan.olle@mpl.mpg.de}
\affiliation{Max Planck Institute for the Science of Light, Staudtstra{\ss}e 2, 91058 Erlangen, Germany}
\author{Remmy Zen}%
\affiliation{Max Planck Institute for the Science of Light, Staudtstra{\ss}e 2, 91058 Erlangen, Germany}
\author{Matteo Puviani}%
\affiliation{Max Planck Institute for the Science of Light, Staudtstra{\ss}e 2, 91058 Erlangen, Germany}
\author{Florian Marquardt}%
\affiliation{Max Planck Institute for the Science of Light, Staudtstra{\ss}e 2, 91058 Erlangen, Germany}
\affiliation{Department of Physics, Friedrich-Alexander Universit\"{a}t Erlangen-N\"{u}rnberg, Staudtstra{\ss}e 5, 91058 Erlangen, Germany}

\date{April 2024}

\begin{abstract}
In the ongoing race towards experimental implementations of quantum error correction (QEC), finding ways to automatically discover codes and encoding strategies tailored to the qubit hardware platform is emerging as a critical problem. Reinforcement learning (RL) has been identified as a promising approach, but so far it has been severely restricted in terms of scalability. In this work, we significantly expand the power of RL approaches to QEC code discovery. Explicitly, we train an RL agent that automatically discovers both QEC codes and their encoding circuits for a given gate set, qubit connectivity and error model, from scratch. This is enabled by a reward based on the Knill-Laflamme conditions and a vectorized Clifford simulator, allowing us to scale our results to 20 physical qubits and distance 5 codes. Moreover, we introduce the concept of a noise-aware meta-agent, which learns to produce encoding strategies simultaneously for a range of noise models, thus leveraging transfer of insights between different situations. Our approach opens the door towards hardware-adapted accelerated  discovery of QEC approaches across the full spectrum of quantum hardware platforms of interest. 
\end{abstract}

\maketitle

\section{Introduction}

% The world is excited about quantum technologies!
There is an ongoing global effort to develop a new generation of quantum technologies with an unprecedented level of control over individual quantum states of many-particle quantum systems. This field encompasses four areas~\cite{Acin_2018}: quantum communication, simulation, computation and sensing, each of them promising to drastically improve on preceding classical technologies. 
An outstanding challenge that is present in all the aforementioned areas is that quantum states are vulnerable to the unwanted effects of noise; if not addressed, the advantages offered over classical technologies disappear altogether.

% QEC is a good strategy to fight noise
\textit{Quantum error correction} (QEC) is a field which emerges from the union of quantum mechanics and classical error correction \cite{Girvin_2023}, and it is the approach that is thought to be essential to achieve maturity in the current wave of quantum technologies. The core idea of QEC is to redundantly embed the quantum information within a subspace (called \textit{code} space) of a \textit{larger} Hilbert space in such a way that different errors map the code space to mutually orthogonal subspaces. If successful, the action of each of these errors can be reverted and the quantum process can continue error-free \citep{inguscio2007proceedings}. 
QEC is \textit{necessary} for large-scale fault-tolerant quantum computing~\citep{preskill1997faulttolerant, Devitt2015}. 
% This field is currently exploding with very cool experiments!
The past few years have witnessed dramatic progress in experimental realizations of QEC on different platforms~\cite{krinner2022realizing,ryan2021realization,postler2022demonstration,cong2022hardware,GoogleQuantum2023}, reaching a point where the lifetime of qubits has been extended by applying QEC~\cite{sivak2023real}.

Since Shor's original breakthrough~\citep{Shor1996},  different qubit-based QEC codes have been constructed, both analytically and numerically, leading to a zoo of codes, each of them conventionally labeled $[[n,k,d]]$, where $n$ is the number of physical qubits, $k$ the number of encoded logical qubits, and $d$ the code distance that defines the number $d-1$  of detectable errors . 
The first examples are provided by the $[[5,1,3]]$ perfect code~\citep{laflamme1996perfect}, the $[[7,1,3]]$ Steane \citep{Steane1996} and the $[[9,1,3]]$ Shor \citep{Shor1996} codes, which encode one logical qubit into 5, 7 and 9 physical qubits, respectively, being able to detect up to 2 physical errors and correct up to 1 error on any physical qubit. 
The most promising approach so far is probably the family of the so-called \textit{toric} or \textit{surface codes}~\cite{kitaev1997quantum}, which encode a logical qubit into the joint entangled state of a $d \times d$ square of physical qubits. 
More recently, examples of quantum Low-Density Parity Check (LDPC) codes that are competitive with the surface code have been discovered~\cite{bravyi2023highthreshold}.

Numerical techniques have already been employed to construct QEC codes. Often, this has involved greedy algorithms, which may lead to sub-optimal solutions but can be relatively fast. For instance, in \citep{Grassl2012} a greedy algorithm was implemented to extend classical linear codes, code concatenation was explored in \citep{Grassl2015concatenation}, and greedy search for finding stabilizer codes was used in \cite{PhysRevA.96.032341}. Often, such numerical methods for QEC code construction were restricted to finding a subclass of codes of a particular structure, e.g. using reduction to classical code search~\cite{10.1063/1.3086833}.

However, knowledge of a code does not automatically translate to knowing how to encode the logical states of that code in an efficient way. The reason is that standard approaches are \textit{unconstrained}, meaning that an all-to-all connectivity between qubits is assumed as well as a set of gates that are not necessarily native to the hardware platform of interest~\cite{gottesman1997stabilizer,Aaronson_2004}. This then leads to larger-than-necessary circuits when implementing them on specific devices.

The recent advent of powerful tools from the domains of machine learning and, more generally, Artificial Intelligence (AI), are transforming the way in which scientific discovery can be achieved~\cite{wang2023scientific}.  
From these, Reinforcement Learning (RL), which is designed to solve complex decision-making problems by autonomously following an action-reward scheme~\citep{sutton1999policy}, is a promising artificial discovery tool for QEC strategies. The task to solve is encoded in a \textit{reward} function, and the aim of RL training algorithms is to maximize such a reward over time. RL can provide new answers to difficult questions, in particular in fields where optimization in a high-dimensional search space plays a crucial role. For this reason, RL can be an efficient tool to tackle the problem of QEC code construction and encoding under hardware-specific constraints.

% RL has ben successfully applied to QEC-related problems before
The first example of RL-based automated discovery of QEC strategies \cite{PhysRevX.8.031084} did not rely on any human knowledge of QEC concepts. While this allowed exploration without any restrictions, e.g. going beyond stabilizer codes, it was limited to only small qubit numbers. More recent works have moved towards optimizing only certain QEC subtasks, injecting substantial human knowledge. For example, RL has been used for optimization of given QEC codes \citep{Nautrup_2019}, and to discover tensor network codes \cite{mauron2023optimization} or codes based on "Quantum Lego" parametrizations \citep{su2023discovery,QuantumLego}.  Additionally, RL has been used to find efficient decoding processes~\cite{andreasson2019quantum,sweke2020reinforcement,colomer2020reinforcement,fitzek2020deep} and self-correcting control protocols \cite{metz2023self}.

% However, there are interesting gaps in using RL for QEC that we aim to close
At this moment, a multitude of experimental platforms are scaling up towards the regime of qubit numbers that make it possible to implement QEC (this includes especially various superconducting qubit architectures, ion traps, quantum dots, and neutral atoms). Given the strong differences in native gate sets, qubit connectivities, and relevant noise models, there is a strong need for a flexible and efficient scheme to automatically discover not only codes but also efficient encoding circuits, adapted to the platform at hand. In our work, we significantly expand the scaling capabilities of RL code discovery by introducing two critical components. The first one is a highly parallelized Clifford circuit simulator that runs entirely on modern AI chip accelerators such as GPUs or TPUs. The second one is an efficient and general reward based on the Knill-Laflamme error correction conditions. The result is a scheme based on deep RL in order to simultaneously discover QEC codes together with the encoding circuit from scratch, tailored to specific noise models, native gate sets and connectivities, minimizing the circuit size for improved hardware efficiency. In particular, our RL agent can be made \textit{noise-aware}, meaning that one and the same agent is able to switch its encoding strategy based on the specific noise that is present in the system. Our approach is flexible and general, and can be readily applied to quantum communication scenarios~\citep{robust_science_2022,repeaters_review}.

While \cite{Cao_2022} also set themselves the task of finding both codes and their encoding circuits, this was done using variational quantum circuits involving continuously parametrized gates, which leads to much more costly numerical simulations and eventually only an approximate QEC scheme. By contrast, our RL-based approach does not rely on any human-provided circuit ansatz, can use directly any given discrete gate set, is able to exploit highly efficient Clifford simulations,  and produces a meta-agent able to cover strategies for a range of noise models.

% The outline of the paper:
The paper is organized as follows: in Section \ref{sec:background} we provide the theoretical background for stabilizer codes, code classification and reinforcement learning. In Section \ref{sec:RLapproach} we describe our approach to build a noise-aware reinforcement learning agent that discovers multiple QEC codes in asymmetric noise channels. We present and analyze our numerical results  in Section \ref{sec:results} and we discuss how our approach can be scaled up to larger code parameters in Section~\ref{sec:towards}.

\section{Background} 
\label{sec:background}
% Conceptual figure illustrating our approach
\begin{figure*}[ht!]
	\includegraphics[width=0.95\textwidth]{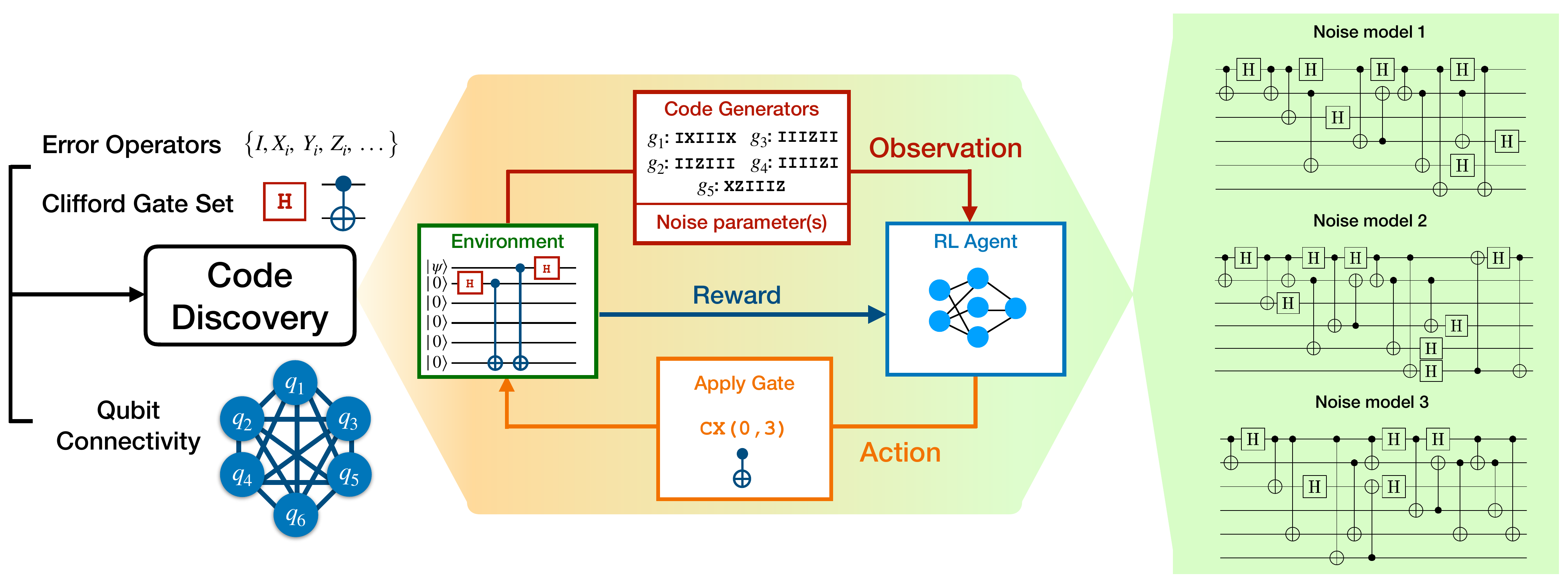}
	\caption{\label{fig:conceptual_figure} QEC code and encoding discovery using a noise-aware RL meta-agent. A set of error operators, a gate set and qubit connectivity are chosen. Different error models can be considered by varying some noise parameters, which are fed as an observation to the agent. The agent then builds a circuit using the available gate set and connectivity that detects the most likely errors from the target error model by using a reward based on the Knill-Laflamme QEC conditions according to Eq.~\eqref{eq:reward}. After training, a single RL agent is able to find suitable encodings for different noise models, which are able to encode \textit{any} state $|\psi\rangle$ of choice.}
\end{figure*}

\subsection{Stabilizer Codes}

Some of the most promising QEC codes are based on the stabilizer formalism \cite{gottesman1997stabilizer}, which leverages the properties of the Pauli group \(G_n\) on \(n\) qubits. The basic idea of the stabilizer formalism is that many quantum states of interest for QEC can be more compactly described by listing the set of \(n\) operators that \textit{stabilize} them, where an operator \(O\) stabilizes a state \(|\psi\rangle\) if \(|\psi\rangle\) is an eigenvector of \(O\) with eigenvalue $+1$: \( O |\psi \rangle = |\psi\rangle\). The Pauli group on a single qubit \(G_1\) is defined as the group that is generated by the Pauli matrices \(X,Y,Z\) under matrix multiplication. Explicitly, $G_1 = \{ \pm I, \pm i I, \pm X, \pm i X, \pm Y, \pm i Y, \pm Z, \pm i Z\}$.  The generalization to \(n\) qubits consists of all \(n\)-fold tensor products of Pauli matrices (called \textit{Pauli strings}). 

A code that encodes \(k\) logical qubits into \(n\) physical qubits is a \(2^k\)-dimensional subspace (the \textit{code space} \(\mathcal{C}\)) of the full $2^n$-dimensional Hilbert space. It is completely specified by the set of Pauli strings $S_\mathcal{C}$ that \textit{stabilize} it, i.e. \(S_\mathcal{C} = \{ s_i \in G_n ~|~ s_i |\psi\rangle = |\psi\rangle,~\forall |\psi\rangle \in \mathcal{C}\}\). $S_\mathcal{C}$ is called the stabilizer group of $\mathcal{C}$ and is usually written in terms of its group generators \(g_i\) as \(S_\mathcal{C} = \langle g_1, g_2, \dots, g_{n-k} \rangle\), where each $g_i$ is a Pauli string.

Noise affecting quantum processes can be represented using the so-called \textit{operator-sum} representation~\cite{nielsen2010quantum}, where a quantum noise channel \(\mathcal{N}\) induces dynamics on the state \(\rho\) according to
\begin{equation}
    \mathcal{N}(\rho) = \sum_\alpha E_\alpha \rho E^\dagger_\alpha~, \label{eq:noise_channel}
\end{equation}
where \(E_\alpha\) are \textit{Kraus operators}, satisfying \(\sum_\alpha E^\dagger_\alpha E_\alpha = I\). 
The most elementary example is the so-called \textit{depolarizing} noise channel,
\begin{equation}
    \mathcal{N}_\text{DP}(\rho) = p_I \rho + p_X X \rho X + p_Y Y \rho Y + p_Z Z \rho Z~,
\end{equation}
where $p_I + p_X + p_Y + p_Z = 1$ and the set of Kraus operators are $E_\alpha = \{ \sqrt{p_I} I, \sqrt{p_X} X, \sqrt{p_Y} Y, \sqrt{p_Z} Z\}$. 
When considering $n$ qubits, one can generalize the depolarizing noise channel by introducing the \textit{global} depolarizing channel,
\begin{equation}
    \mathcal{N}_\text{GDP}(\rho) = \bigotimes_{j=1}^n \mathcal{N}^{(j)}_\text{DP}(\rho_j)~, \label{eq:global_depolarizing_noise_model}
\end{equation}
consisting of local depolarizing channels acting on each qubit $j$ independently. 
Taken as is, this error model generates \textit{all} \(4^n\) Pauli strings by expanding \eqref{eq:global_depolarizing_noise_model}. A commonly used simplification is the following. Assume that all error probabilities are identical, i.e. $p_X =  p_Y = p_Z \equiv p$ (and $p_I = 1-3p$). Then, the probability that a given error occurs decreases with the number of qubits it affects. For instance, if we consider 3 qubits, the probability associated with $XII$ is $p(XII) = p (1-3p)^2$, and in general the leading order contribution to the probability of an error affecting $m$ qubits is $p^m$. 
This leads to the concept of the \textit{weight} of an operator as the number of qubits on which it differs from the identity and to a hierarchical approach to building QEC codes. In particular, stabilizer codes are described by specifying what is the minimal weight in the Pauli group that they cannot detect. 

The fundamental theorem in QEC is a set of necessary and sufficient conditions for quantum error detection discovered independently by Bennett, DiVincenzo, Smolin and Wootters~\cite{BDW96}, and by Knill and Laflamme in \cite{PhysRevA.55.900} (KL conditions from now on). These state that a code \(\mathcal{C}\) with associated stabilizer group \(S_\mathcal{C}\) can \textit{detect} a set of errors \(\{ E_\mu\} \subseteq G_n\), if and only if for all $E_\mu$ we have either
\begin{equation}
    \{ E_\mu , g_i\} = 0~, \label{eq:KL_anticommute}
\end{equation}
for at least one \(g_i\), or the error itself is harmless, i.e.
\begin{equation}
    E_\mu \in S_\mathcal{C}~. \label{eq:KL_in_S}
\end{equation}
The  smallest weight in \(G_n\) for which none of the above two conditions hold is called the \textit{distance} of the code. For instance, a distance$-3$ code is capable of detecting \textit{all} Pauli strings of up to weight 2, meaning that KL conditions \eqref{eq:KL_anticommute}, \eqref{eq:KL_in_S} are satisfied for \textit{all} Pauli strings of weights 0, 1 and 2. Moreover, the \textit{smallest} weight for which these are not satisfied is 3, meaning that there is \textit{at least one} weight$-3$ Pauli string violating both \eqref{eq:KL_anticommute} and \eqref{eq:KL_in_S}. However, \textit{some} weight$-3$ Pauli strings (and higher weights) will satisfy the KL conditions, in general.

While these conditions are framed in the context of quantum error detection, there is a direct correspondence with quantum error correction. Indeed, a quantum code of distance $d$ can \textit{correct} all errors of up to weight \(t = \lfloor(d-1)/2 \rfloor\)~\cite{gottesman1997stabilizer}. 
If all the errors that are detected with a weight smaller than \(d\) obey (\ref{eq:KL_anticommute}), the code is called \textit{non-degenerate}. On the other hand, if some of the errors satisfy (\ref{eq:KL_in_S}), the code is called \textit{degenerate}.

The default weight-based $[[n,k,d]]$ classification of QEC codes implicitly assumes that the error channel is symmetric, meaning that the probabilities of Pauli X, Y and Z errors are equal. However, this is usually not the case in experimental setups: for example, dephasing ($Z$ errors) may dominate bit-flip (X) errors.
In our work, we will consider an asymmetric noise channel where  \(p_X = p_Y\) but $p_X \neq p_Z$. To quantify the asymmetry, we use the bias parameter \(c_Z\) \cite{Cao_2022}, defined as
\begin{equation}
    c_Z = \frac{\log p_Z}{\log p_X}~.
    \label{eq:bias-parameter}
\end{equation} 
For symmetric error channels, $c_Z = 1$. If Z-errors dominate, then $0 < c_Z < 1$, since $p_Z=p_X^{c_Z}$ and $p_X, p_Z\ll1$; conversely $c_Z > 1$ when X/Y errors are more likely than Z errors.
The weight of operators and the code distance can both be generalized to asymmetric noise channels  \cite{PhysRevA.75.032345, Asymmetric2010, Asymmetric2011, LaGuardia2014}. Consider a Pauli string operator \(E_\mu\) and denote as $w_X$ the number of Pauli $X$ inside $E_\mu$ (likewise for $Y$, $Z$). Then one can introduce the \(c_Z-\)effective weight~\cite{Cao_2022} of $E_\mu$ as
\begin{equation}
    w_e(E_\mu, c_Z) = w_X(E_\mu) + w_Y(E_\mu) + c_Z w_Z(E_\mu)~,
    \label{eq:effective-weight}
\end{equation}
which reduces to the symmetric weight for $c_Z=1$, as expected. The \(c_Z-\)effective distance of a code \(d_e(c_Z)\) is then defined~\cite{Cao_2022} as the largest possible integer such that the  KL conditions \eqref{eq:KL_anticommute}, \eqref{eq:KL_in_S}  hold for all Pauli strings $E_\mu$ with $w_e(E_\mu,c_Z)<d_e(c_Z)$. Like in the symmetric noise case, the meaning of this effective distance is that \textit{all} error operators with an effective weight smaller than $d_e$ can be detected.

\subsection{Code Classification}

It is well known that there is no unique way to describe quantum codes. For instance, there are multiple sets of code generators that generate the same stabilizer group, hence describing the \textit{same} code. Moreover, the choice of logical basis is not unique and qubit labeling is arbitrary. While such redundancies are convenient for describing quantum codes in a compact way, comparing and classifying different codes can be rather subtle. Fortunately, precise notions of code equivalence have been available in the literature since the early days of this field. In this work, we will refer to \textit{families} of codes based on their quantum weight enumerators (QWE)  \cite{shor1996quantum}, \(A(z)\) and \(B(z)\), which are polynomials with coefficients
\begin{eqnarray}
    A_j &=& \frac{1}{(2^k)^2} \sum_{w(E_\mu)=j} \Tr \left(E_\mu P_\mathcal{C}\right) \Tr \left( E^\dagger_\mu P_\mathcal{C}\right)~, \nonumber\\
    B_j &=& \frac{1}{2^k} \sum_{w(E_\mu)=j} \Tr \left( E_\mu P_\mathcal{C} E^\dagger_\mu P_\mathcal{C} \right)~,
    \label{eq:quantum-weight-enumerators}
\end{eqnarray}
where $w$ is the operator ($c_Z=1$) weight, \(j\) runs from \(0\) to \(n\) and $P_{\mathcal C}$ is the orthogonal projector onto the code space. Intuitively, \(A_j\) counts the number of error operators of weight \(j\) in \(S_\mathcal{C}\) while \(B_j\) counts the number of error operators of weight \(j\) that commute with all elements of \(S_\mathcal{C}\). Logical errors are thus the ones that commute with \(S_\mathcal{C}\) but are not in \(S_\mathcal{C}\), and these are counted with $B_j - A_j$.

Such a classification is especially useful in scenarios with symmetric noise channels, where it is irrelevant whether the undetected errors contain a specific Pauli operator at a specific position. However, such a distinction can in principle be important in asymmetric noise channels. One could in principle generalize \eqref{eq:quantum-weight-enumerators} to asymmetric noise channels substituting the weight $w$ by the effective weight $w_e$ of operators, but then comparing codes across different values of noise bias becomes cumbersome. Hence, in the present work we will always refer to (symmetric) code families according to \eqref{eq:quantum-weight-enumerators} for all values of $c_Z$, i.e. we will effectively pretend that $c_Z=1$ when computing the weight enumerators of asymmetric codes.

\subsection{Reinforcement Learning}
Reinforcement Learning (RL)~\cite{sutton2018reinforcement} is designed to discover optimal action sequences in decision-making problems. The goal in any RL task is encoded by choosing a suitable \textit{reward} \(r\), a quantity that measures how well the task has been solved, and consists of an \textit{agent} (the entity making the decisions) interacting with an \textit{environment} (the physical system). In each time step \(t\), the environment’s state \(s_t\) is observed. Based on this observation, the agent takes an action \(a_t\) which then affects the current state of the environment. A \textit{trajectory} is a sequence of state and action pairs that the agent takes. An \textit{episode} is a trajectory from an initial state to a terminal state. For each action, the agent receives a reward \(r_t\), and the goal of RL algorithms is to maximize the expected cumulative reward (return), \( \mathbb{E} \left[ \sum_t r_t \right]\). The agent’s behavior is defined by the \textit{policy} \(\pi_\theta (a_t|s_t) \), which denotes the probability of choosing action \(a_t\) given observation \(s_t\), and that we parameterize by a neural network with parameters \(\theta\).
Within RL, policy gradient methods~\cite{sutton1999policy} optimize the policy by maximizing the expected return with respect to the parameters \(\theta\) with gradient ascent. One of the most successful algorithms within policy gradient methods is the actor-critic algorithm~\cite{konda1999actor}. The idea is to have two neural networks: an actor network that acts as the agent and that defines the policy, and a critic network, which measures how good was the action taken by the agent. In this paper, we use a state-of-the-art policy-gradient actor-critic method called Proximal Policy Optimization (PPO)~\cite{schulman2017proximal}, which improves the efficiency and stability of policy gradient methods.

\section{Reinforcement Learning Approach to QEC Code Discovery} \label{sec:RLapproach}

\begin{figure*}[ht!]
    \centering
    \includegraphics[width=0.97\textwidth]{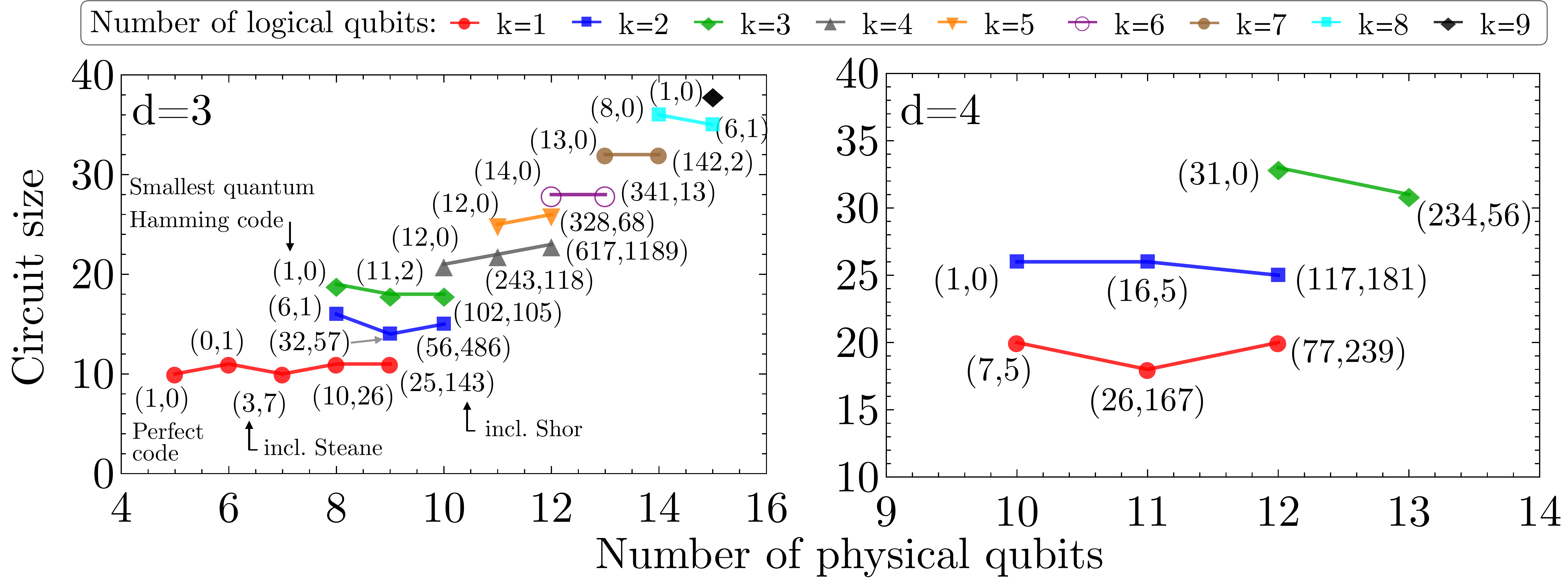}
    \caption{Discovering codes and encoding circuits for various numbers of physical qubits, logical qubits, and distances (see main text and Appendix~\ref{appendix:distance_5} for $d=5$). Families of stabilizer codes tailored to symmetric depolarizing noise channels, found with our RL framework. The labels \((x,y)\) indicate the number of non-degenerate $(x)$ and degenerate $(y)$ code families. The circuit size shown is the absolute minimum throughout all families, and different families in general have different minimal circuit sizes. Since further training runs do not increase family populations, it is likely that there are no more stabilizer codes for the shown code parameters.}
    \label{fig:results_families}
\end{figure*}

The main objective of this work is to automatize the discovery of QEC codes and their encoding circuits using RL. We will consider a scenario where the encoding circuit is assumed to be error-free (non fault-tolerant encoding). This is applicable to quantum communication or quantum memories, where the majority of errors happen during transmission over a noisy channel or during the time the memory is retaining the information.
Eventually we will show that it is possible to train a meta-agent that is capable of adapting its strategy according to the noise model, without any retraining. This leverages the concept of transfer learning, where improvements gained in training for one scenario (here, one value of a noise parameter) carry over and accelerate the training progress for other scenarios. A scheme of our approach can be found in Fig.~\ref{fig:conceptual_figure}.

\subsection{Encoding Circuit}
% Ab initio/tabula rasa approach: from an empty circuit to a code
In order to encode the state of \(k\) logical qubits on \(n\) physical qubits one must find a sequence of quantum gates that will entangle the quantum information in such a way that QEC is possible with respect to a target noise channel. Initially, we imagine the first \(k\) qubits as the original containers of our (yet unencoded) quantum information, which can be in any state $|\psi\rangle \in (\mathbb{C}_2)^{\otimes k}$. The remaining $n-k$ qubits are chosen to each be initialized in the state $|0\rangle$. These will be turned into the corresponding logical state $|\psi\rangle_L \in (\mathbb{C}_2)^{\otimes n}$ via the application of a sequence of Clifford gates on any of the $n$ qubits (where Clifford gates are defined to be those that map Pauli strings to Pauli strings and are generated by the Hadamard \(H\), the Phase \(S\) and the CNOT gates). In the stabilizer formalism, this means that initially the generators of the stabilizer group are
\begin{equation}
    Z_{k+1},~Z_{k+2}, \dots, Z_n~. \label{eq:initial_generators}
\end{equation}
The task of the RL agent is to discover a suitable  encoding sequence of gates for the particular error model under consideration. After applying each gate, the $n-k$ code generators \eqref{eq:initial_generators} are updated. The agent then receives a representation of these generators as input (as its observation) and suggests the next gate (action) to apply.  In this way, an encoding circuit is built up step by step, taking into account the available gate set and connectivity for the particular hardware platform. This process terminates when the KL conditions (\ref{eq:KL_anticommute}), (\ref{eq:KL_in_S}) are satisfied for the target error channel and the learned circuit can then be used to encode \textit{any} state $|\psi\rangle$ of choice. To gain clarity in our exposition, we include in Appendix~\ref{appendix:three_qubit_repetition_code_example} a detailed description of the encoding process of the three-qubit repetition code.

\subsection{Reward}
The most delicate matter in RL problems is building a suitable reward for the task at hand. Our goal is to design an agent that, given an error model that includes a set of errors $\{E_\mu\} \subseteq G_n$ with associated occurrence probabilities $\{p_\mu\}$, is able to find an encoding sequence that protects the quantum information from such noise. 
% Though experiment with Alice and Bob: What would they want to optimize?
Let us consider a quantum communication setup to have a concrete picture. Here, we imagine Alice and Bob exchanging some quantum bits of information contained in a state $|\psi\rangle$. We assume that they are able to implement all gates and measurements without errors and that errors only happen while the message is traveling through the communication channel, which they have previously characterized. Alice encodes state $|\psi\rangle$ into $|\psi\rangle_L$ and sends it through the noisy channel, after which it is received by Bob in the form of a possibly corrupted, mixed state. Since Bob knows the encoding that Alice has used, he also knows what are the stabilizer generators. Hence, Bob proceeds to perform syndrome measurements according to the stabilizer generators of the code and finally corrects the errors that he believes have happened along the way. If done perfectly, the corrected state  becomes the original state $|\psi\rangle_L$ sent by Alice. However, even when both Alice and Bob have access to perfect gates and measurements, the probability of recovering the state is not one. The reason is that multiple errors may trigger the same syndrome measurement, and in some cases Bob will mistakenly correct for an error that did not actually happen. An option for our RL agent could thus be to maximize the probability of recovering the initial state, or what is the same, minimizing the probability that Bob applies a wrong error correction operation. Unfortunately, optimizing for this task is computationally very expensive. Indeed, for each syndrome, there are different errors that could have triggered it. On top of that, from all the errors that could have triggered that syndrome, computational resources have to be employed in finding the most dangerous one, which then has to be selected as the candidate error to be corrected. Such expensive calculations would have to be carried out in every step of the RL procedure.
% The ideal optimization is prohibitively expensive, but we have an alternative
A much cheaper alternative that avoids performing such error categorization is to use a scheme where the cumulative reward (which RL optimizes) simply is maximized whenever all the KL conditions are fulfilled.  One implementation of this idea uses the (negative) weighted KL sum as an instantaneous reward:
\begin{equation}
    r_t = - \sum_\mu \lambda_\mu K_\mu~, \label{eq:reward}
\end{equation}
where $K_\mu = 0$ if either (\ref{eq:KL_anticommute}) or (\ref{eq:KL_in_S}) are satisfied for the corresponding error operator $E_\mu$, and $K_\mu = 1$ otherwise. Here $\lambda_\mu$ are real positive numbers that we will take to be hyperparameters quantifying how dangerous each error is. For any choice of $\lambda_\mu$, if all errors in \( \{ E_\mu \} \) can be detected, the reward is zero, and is negative otherwise, thus leading the agent towards the goal we set out to solve. The range of the index $\mu$ is found by counting the number of Pauli strings of weight $w < d$, which is
\begin{equation}
    \left| \{E_\mu\}\right|_{w < d} = \sum_{w=0}^{d-1} 3^w \binom{n}{w}~, \label{eq:numE}
\end{equation}
where the factor of three is for $X,Y,Z$. Thus, the fact that \eqref{eq:numE} grows exponentially with \(d\) will impose the most severe limitation in our approach (as is the case in \textit{any} QEC application). 
% The hyperparameters are related to the error probabilities
Later, we will also be interested in situations where not all errors can be corrected simultaneously and a good compromise has to be found. In that case, one simple heuristic choice for the reward \eqref{eq:reward} would be $\lambda_\mu = p_\mu$, giving more weight to errors that occur more frequently. While we will later see that maximizing the KL reward given here is not precisely equivalent to minimizing the overall probability of incorrectly classifying errors, one can still expect a reasonable performance at this task, which would be the ultimate goal in any QEC scheme.

% Our reward is nice: it gives us flexibility and favors small circuits
A nice feature of our reward is that one can favor certain types of codes. For instance, we can target non-degenerate codes only by ignoring (\ref{eq:KL_in_S}). Moreover, making the reward non-positive favors short gate sequences, which is desirable when preparing these codes in actual quantum devices. 

As a final remark, we note that RL optimizes the cumulative reward $R$, i.e. the sum of $r_t$ over all time steps. Therefore, defining the instantaneous reward $r_t$ in terms of the weighted KL sum means we are asking the agent that it keeps this sum as small as possible \textit{on average}.  While this might seem less desirable than setting $R$ itself to be the KL sum at the final time step, we have heuristically found that the ansatz adopted here produces more stable and faster training behavior than by taking $R$ itself to be the KL sum.

\subsection{Noise-aware meta-agent}

Regarding the error channel to be targeted, here there are in principle several choices that can be made. The most straightforward one is choosing a global depolarizing channel as given by \eqref{eq:global_depolarizing_noise_model}. This still allows for asymmetric noise, i.e. different probabilities $p_X, p_Y, p_Z$. One option would be to train an agent for any given, fixed choice of these probabilities, necessitating retraining if these characteristics change. However, we want to go beyond that and discover a single agent being capable of deciding what is the optimal encoding strategy for \textit{any} level of bias in the noise channel \eqref{eq:bias-parameter}. For instance, we want this noise-aware agent to be able to understand that it should prioritize detecting more $Z$ errors than $X$ ones when the channel is biased towards $Z$, yet it should do the opposite when $X$ errors become more likely.
This translates into two aspects: The first one is that the agent has to receive the noise parameters as input. In the illustrative example further below, we will choose to supply the bias parameter $c_Z$ as an extra observation, while keeping the overall error probability fixed. The second aspect is that the list of error operators will have to contain more operators than the total number that can actually be detected reliably, since it is now part of the agent's task to prioritize some of those errors while ignoring the least likely errors. All in all, the list of operators participating in the reward \eqref{eq:reward} will be fixed and we will choose those with at most (symmetric) weight $d-1$, for a certain $d$; the idea is then that when varying $c_Z$, the probabilities of these errors will change and the agent will have to figure out which are the most dangerous errors that must be detected in every case.

\subsection{Vectorized Clifford simulator}

We finally comment on a few more detailed aspects of our implementation, with further details in Appendix~\ref{appendix:simulations}. 
RL algorithms exploit trial-and-error loops until a signal of a good strategy is picked up and convergence is reached, so it is of paramount importance that simulations of our RL environment are extremely fast. Luckily, thanks to the Gottesman-Knill theorem, the Clifford circuits needed here can be simulated efficiently on classical computers. Optimized numerical implementations exist, e.g. \textsc{Stim} \cite{gidney2021stim}. However, in an RL application we want to be able to run multiple circuits in parallel in an efficient, vectorized way that is compatible with modern machine learning frameworks. For that reason, we have implemented our own special-purpose vectorized Clifford simulator, which is publicly available in our repository~\cite{olle2024qdxgithub}. Briefly, we use the symplectic binary formalism of the Pauli group~\cite{Aaronson_2004} to represent the stabilizer generators. $S_\mathcal{C}$ is then represented by a \textit{check matrix} $H$~\cite{Aaronson_2004}, which is a $(n-k) \times 2n$ binary matrix where each row $i$ represents the Pauli string $g_i$ from $S_\mathcal{C}$. Clifford gates are also implemented using binary matrices, and we achieve a massive parallelization by running Clifford-based simulations of quantum circuits in parallel, meaning that the agent interacts with a batch of RL environments (quantum circuits) at every timestep. Our Clifford simulator is implemented using \textsc{Jax}~\cite{jax2018github}, a state-of-the-art modern machine learning framework with good vectorization and just-in-time compilation capabilities. On top of that, we also train multiple RL agents in parallel on a single GPU. This is achieved by interfacing with \textsc{PureJaxRL} \cite{lu2022discovered}, a library that offers a high-performance end-to-end \textsc{Jax} RL implementation. We have included further implementation details and performance of the RL algorithm used in this work in Appendix~\ref{appendix:hyperparameters}. The source code for our project is available on \textsc{GitHub} under the name QDX~\cite{olle2024qdxgithub}, which is an acronym for Quantum Discovery with \textsc{Jax}. It includes both the Clifford simulator, the PPO algorithm and demo Jupyter notebooks to reproduce some of our main results.

The efficiency of Clifford simulations, as well as the use of powerful state-of-the-art RL algorithms, enables us to easily go beyond the recent results of \cite{Cao_2022}, which were based on variational quantum circuits. In particular, we are able to straightforwardly discover codes and encoding circuits for both larger number of qubits (14 vs 20) and larger code distances (4 vs 5).

\section{Results} 
\label{sec:results}

We will first illustrate the basic workings of our approach for a symmetric noise channel before introducing the meta-agent that is able to simultaneously discover strategies for a range of  noise models.

\subsection{Codes in a symmetric depolarizing noise channel}
\begin{figure}[ht!]
    \centering
    \includegraphics[width=0.4\textwidth]{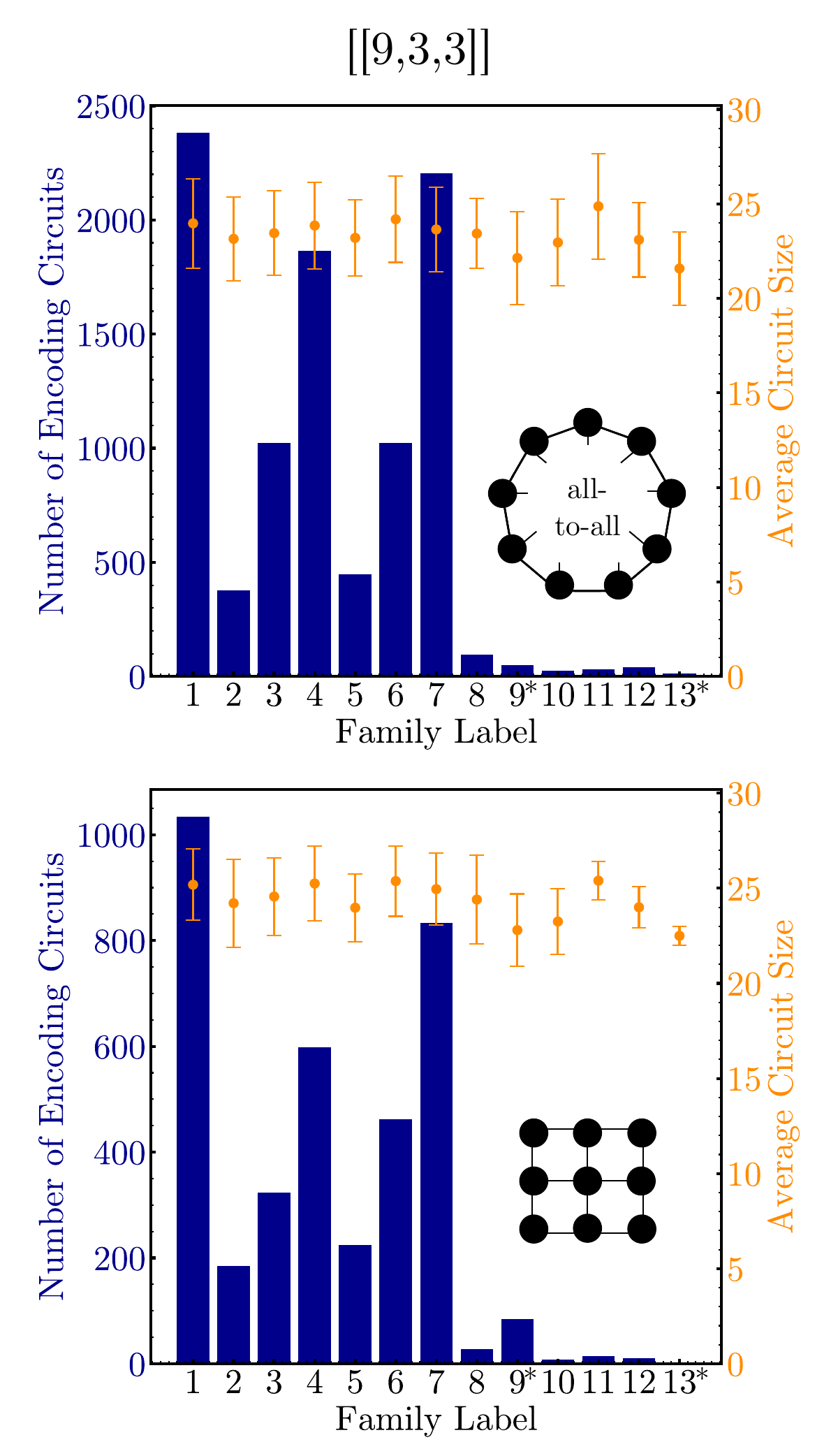}
    \caption{Influence of connectivity. Characteristics of the 13 families of $\code{9,3,3}$ codes found with our framework, clustered according to families distinguished by their quantum weight enumerators \eqref{eq:quantum-weight-enumerators}. Families 9 and 13 (*) are degenerate, while the rest are non-degenerate. We have trained a total of 10240 agents for each of both cases. In the all-to-all (directed: $\text{CNOT}(i<j)$) connectivity, 9574 agents were successful, while this number went down to 3808 in the other case. The bars display how these codes are distributed across different families. Codes in the same family found by different agents are not necessarily distinct, so the bars are rather an indication of the likelihood of a training run to find a code within the family. The points show the mean circuit size, averaged within each family, while the error bar is its standard deviation. It is interesting to see that even with different connectivities, families occur with similar likelihoods during training. We explicitly list the corresponding quantum weight enumerators computed with \eqref{eq:quantum-weight-enumerators} in Appendix~\ref{appendix:QWE_9_3_3}.}
    \label{fig:histogram_9_3_3}
\end{figure}

We now illustrate the versatility of our approach by discovering a library of different \(\code{n,k,d}\) codes and their associated encoding circuits. 

We fix the error model to be a symmetric depolarizing channel with error probability $p$, meaning $p_I = 1-3p$, $p_X = p_Y = p_Z = p$, and thus no noise parameter is needed. We also vary the target code distance \(d\) from 3 to 5. The corresponding target error set is $E_\mu = \{I, X_i, Y_i, Y_j, X_i X_j, \dots , Z_i Z_j\}$ for $d=3$, and likewise for $d=4, 5$, with the set for $d=5$ including all Pauli string operators of up to weight 4. In our numerical experiments, we choose $p_I = 0.9$, which enters the reward function where we choose the weights proportional to the error probabilities (recall the discussion above). Although the final discovered codes and encoding circuits are independent of this choice, it can affect the learning progress.

For illustrative purposes, we take the gateset to be \(\{H_i, \text{CNOT}(i<j)\}\), i.e. a \textit{directed} all-to-all connectivity, which is sufficient given that our unencoded logical state is at the first qubits by design. Nevertheless, we will also see examples with alternative gatesets and other connectivities. 
The final physical hyperparameter is the maximal number of gates that we allow before restarting the learning trajectory. This number will be varied from 20 to 50, depending on the target code parameters. Unless we say so explicitly, we target both non-degenerate and degenerate codes.

\begin{figure*}[ht!]
    \centering
    \includegraphics[width=0.9\textwidth]{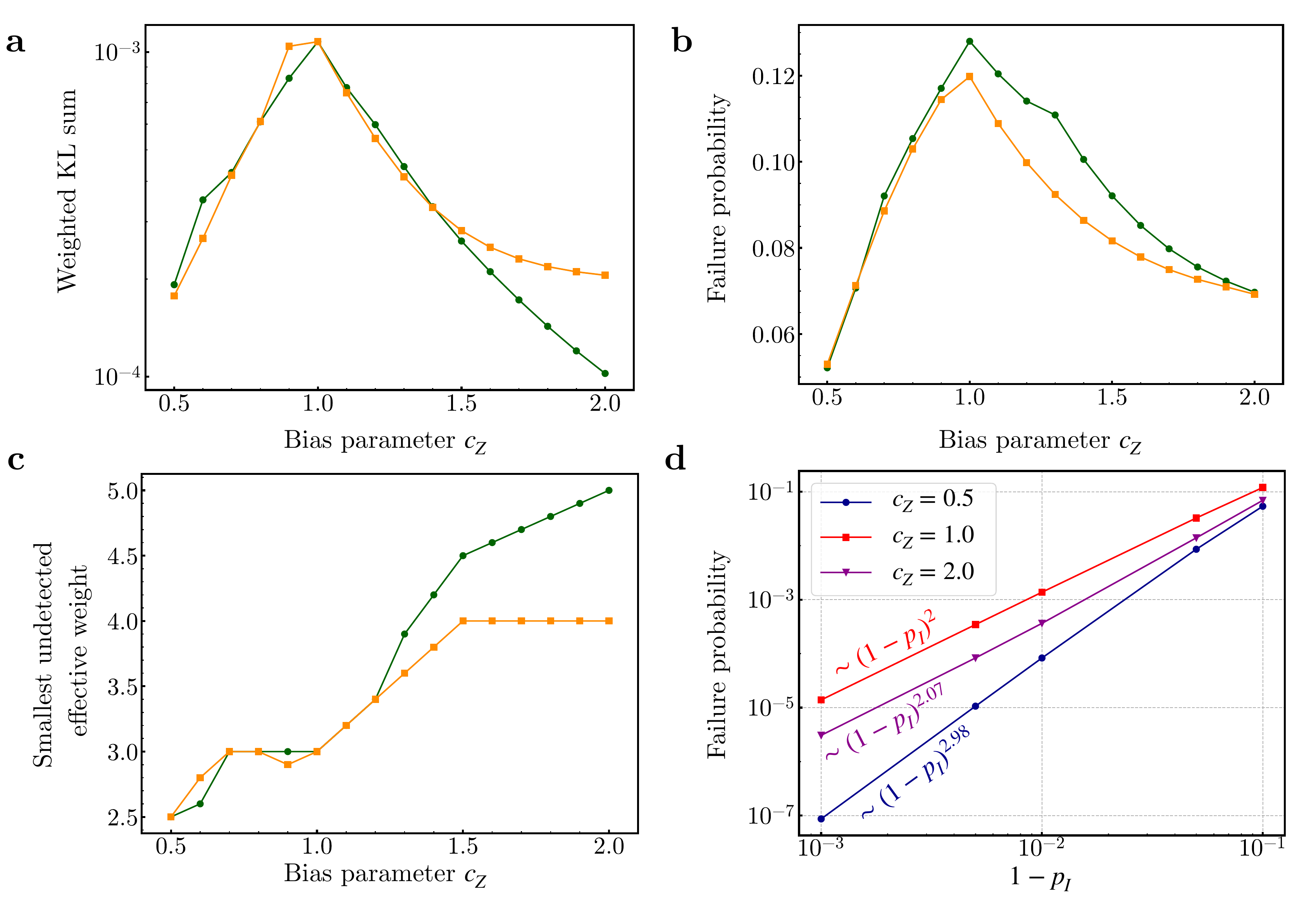}
    \caption{Performance of the noise-aware RL agent. The agent finds $n=9, k=1$ codes and encoding circuits, simultaneously for different levels of noise bias $c_Z$, with single-qubit fidelity $p_I=0.9$. In panels a,b,c, green represents the agent that was post-selected among all trained agents for performing best at minimizing the weighted KL sum, averaged over all $c_Z$ values. Orange refers to the agent minimizing the failure probability, averaged over $c_Z$. \textbf{a} Weighted KL sum as a function of the noise bias parameter $c_Z$  (best agent: green line). \textbf{b} Failure probability as a function of the noise bias parameter $c_Z$ (best agent: orange line) \textbf{c} Smallest undetected effective weight (effective code distance is the integer part) as a function of the noise bias parameter $c_Z$. While there is almost a perfect overlap between both best agents until $c_Z=1.1$, the situation changes afterwards, leading at $c_Z=2$ to a $d_e=5$ code (green) or a $d_e=4$ code (orange) that perform equally well in terms of the failure probability, as seen in \textbf{b}. \textbf{d} Evaluation of the failure-probability of the best-performing agent (orange in the other panels) for larger values of $p_I$ (smaller errors) than the ones it was trained on.}
    \label{fig:noise_aware_result_9}
\end{figure*}

For $d=3$ and $d=4$ codes we proceed as follows: for any given target \(\code{n,k,d}\), we launch a few training runs. Once the codes are collected, we categorize them by calculating their quantum weight enumerators, see Eq.~(\ref{eq:quantum-weight-enumerators}), leading to a certain number of non-degenerate ($A_{d-1} = B_{d-1} = 0$) and degenerate ($A_{d-1} = B_{d-1} \neq 0$) families. We repeat this process and keep launching new training runs until no new families are found by further runs. In this way, our strategy presumably finds \textit{all} stabilizer codes that are possible for the given parameters $n,k,d$. This total number of families is shown in Fig.~\ref{fig:results_families}, with labels $(x,y)$ for each $\code{n,k,d}$, where $x$ is the number of non-degenerate families and $y$ is the number of degenerate ones. It should be stressed that categorizing all stabilizer code families is in general an NP-complete problem~\cite{yu2007graphical}, yet our framework is very effective at solving this task. To the best of our knowledge, this work provides the most detailed tabulation of $(x,y)$ populations together with optimal encoding circuits for the code parameters shown here.

This approach discovers suitable encoding circuits, given the assumed gate set, for a large set of codes. Among them are the following known codes for $d=3$ (see ~\cite{yu2013all} for explicit constructions of codes $\code{n,n-r,3}$ with minimal $r$, for all $n$): The first one is the five-qubit perfect code \cite{laflamme1996perfect}, which consists of a \textit{single} non-degenerate $\code{5,1,3}$ code family and is the smallest stabilizer code that corrects an arbitrary single-qubit error. Next are the 10 families \cite{yu2007graphical} of $\code{7,1,3}$ codes, one of which corresponds to Steane's code~\cite{Steane1996}. 
The smallest single-error-correcting surface code, Shor's code~\cite{Shor1996}, is rediscovered as one of the 143 degenerate code families with parameters $\code{9,1,3}$. The smallest quantum Hamming code~\cite{PhysRevA.54.1862} $\code{8,3,3}$ is obtained as well. Our approach is efficient enough to reach up to 20 physical qubits. The largest code parameters that we have considered for $d=3$ is $\code{20,13,3}$, finding an encoding circuit with 45 gates (see Appendix~\ref{appendix:encoding_circuits}).

The circuit size shown for each $\code{n,k,d}$ in Fig.~\ref{fig:results_families} is the minimal one found across all discovered families. In general, different families have different circuit sizes, and even within the same family we find variations in circuit sizes. 

The RL framework presented here easily allows to find encoding circuits for different connectivities. The connectivity affects the  likelihood of discovering codes within a certain family during RL training as well as the typical circuit sizes.  In Fig.~\ref{fig:histogram_9_3_3} we illustrate this for the case of  $\code{9,3,3}$ codes, with their 13 families, for two different connectivities: an all-to-all (directed, i.e. $\text{CNOT}(i<j)$) and a nearest-neighbor square lattice connectivity (see also Appendix~\ref{appendix:connectivity_and_gateset} for examples using different gatesets and a larger variety of connectivities). More precisely, we train 10240 agents for each of both cases and find that 9574 of these were successful in the all-to-all connectivity, while this number went down to 3808 for the square lattice connectivity, given a finite allotted time. The frequency with which each code family was found can be seen to be comparably similar in both cases with the sole exception of family 9 (degenerate), which is found more frequently using the square lattice connectivity. On average, the agent needs one less gate to prepare the encoding on the all-to-all connectivity than when having using the square lattice. 

We now move to distance $d=5$ codes. These are more challenging to find due to the significantly increased number of error operators (\ref{eq:numE}) to keep track of, which impacts both the computation time and the hardness of satisfying all KL conditions simultaneously. Nevertheless, our strategy is also successful in this case. It is known that the smallest possible distance$-5$ code has parameters \(\code{11,1,5}\), a result that we confirm with our strategy. We find the single family of this code to have weight enumerators,
\begin{align}
A &= (1, 0, 0, 0, 0, 0, 198, 0, 495, 0, 330, 0)~, \\
B &= (1, 0, 0, 0, 0, 198, 198, 990, 495, 1650, 330, 234) ~, \nonumber
\end{align}
with an encoding circuit consisting of 32 gates in the minimal example, which we show in Appendix~\ref{appendix:encoding_circuits}.

The largest $d=5$ code that we have considered is $\code{15,2,5}$. Due to the larger code parameter values, we have restricted the search to non-degenerate codes. We have found a single code family with weight enumerators
\begin{align}
A &= (1, 0, 0, 0, 0, 0, 23, 96, 361, 776, 1318, 1832, \nonumber \\
 &~ \quad 1814, 1304, 579, 88) ~, \nonumber \\
B &= (1, 0, 0, 0, 0, 101, 449, 1763, 5081, 12034, \nonumber \\ 
 &~ \quad 21722, 29366, 29622, 20489, 8661, 1783) ~.
\end{align}
and an encoding circuit consisting of 49 gates shown in Appendix~\ref{appendix:encoding_circuits}. Other code parameters for $d=5$ that we have successfully discovered are shown in Appendix~\ref{appendix:distance_5}.

\subsection{Noise-aware meta-agent}
\begin{figure*}[ht!]
    \centering
    \includegraphics[width=0.9\textwidth]{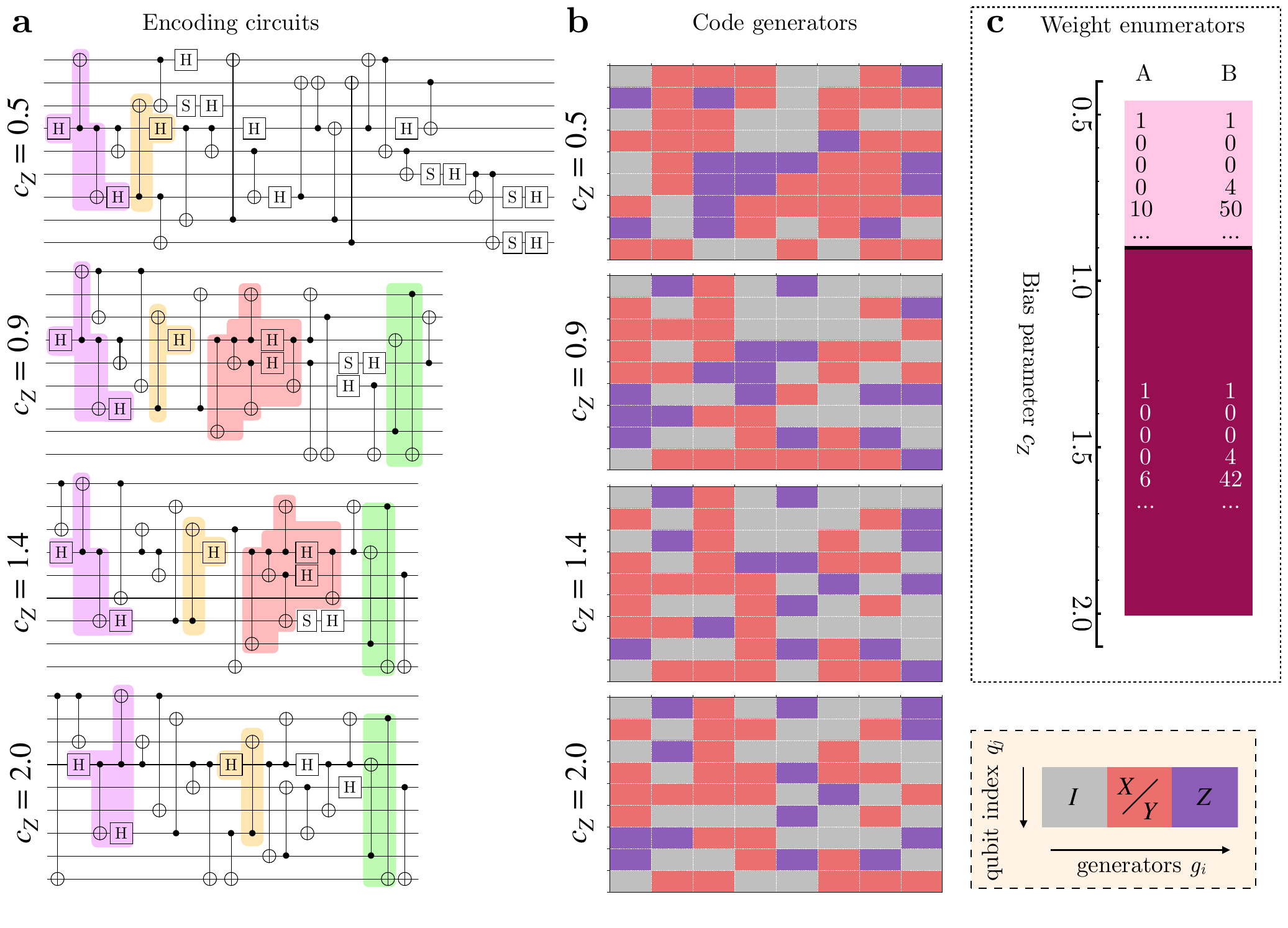}
    \caption{Characteristics of the 9-qubit codes and encodings found by the noise-aware meta-agent post-selected for minimizing the failure probability. \textbf{a} Encoding circuits: Here we see that many small gate sequences (highlighted with different colors) are reused across different values of $c_Z$. This is an indication of transfer learning, i.e. the power of the meta-agent. \textbf{b} Code generators $g_i$ corresponding to the encoding circuits, where we do not make a distinction between $X$ or $Y$. Here we see that the code generators $g_i$ vary across different values of $c_Z$. \textbf{c} Associated code family according to their (symmetric) weight enumerators $A$, $B$. The same code family is used from $0.5 \leq c_Z < 0.9$, while a family switching occurs at $c_Z=0.9$, and it is kept until $c_Z=2$.}
    \label{fig:noise_aware_circuits}
\end{figure*}

We now move on to codes in more general asymmetric depolarizing noise channels. This lets us illustrate a powerful aspect of RL-based encoding and code discovery: One and the same agent can learn to switch its encoding strategy depending on some parameter characterizing the noise channel. This can be realized by training this noise-aware agent on many different runs with varying choices of the parameter, which is fed as an additional input to the agent. One motivation for this approach is that the agent may learn to generalize, i.e. transfer what it has learned between different values of the parameter. 

In the present example, the parameter in question is the bias parameter  \(c_Z = \log p_Z / \log p_X\) introduced above, Eq.~\eqref{eq:bias-parameter}. This allows the \textit{same} agent to switch its strategy depending on the kind of bias present in the noise channel. Once a particular value of $c_Z$ is chosen, the error probabilities characterizing the noise channel are $(p_I, p_X, p_X, p_X^{c_Z})$. Normalization of the error probabilities imposes a relationship between $p_I$ and $p_X$, which means that there is only one other free parameter besides $c_Z$, either $p_I$ or $p_X$. It is more beneficial for training and generalization to keep $p_I$ fixed and solve for $p_X$; otherwise the magnitude of the probabilities $\{p_\mu\}$ changes a lot when varying $c_Z$, leading to poorer performance.

The error set $E_\mu$ is now taken to be all Pauli strings of (symmetric) weight $\leq 4$, i.e. $\{E_\mu\} = \{ I, X_i, Y_i, Z_i, X_i X_j, \dots,  Z_i Z_j Z_k Z_l\}$, but their associated error probabilities (and thus their effective weights according to (\ref{eq:effective-weight})) will vary depending on $c_Z$. For every RL training trajectory, a new $c_Z$ is chosen and the error probabilities $p_\mu$ are updated correspondingly. For the number of physical qubits that we will consider, the KL conditions \eqref{eq:KL_anticommute}, \eqref{eq:KL_in_S} cannot be exactly satisfied. Hence, we are forcing the agent to achieve some compromise: the most likely errors will have to be detected at the expense of not detecting other, less likely ones. 

Regarding more detailed aspects of our implementation, we sample $c_Z$ from the set $c_Z \in \{ 0.5, 0.6, 0.7, \dots , 1.9, 2\}$ with a uniform probability distribution. The hyperparameters $\lambda_\mu$ of the reward \eqref{eq:reward} are defined as
\begin{equation}
    \lambda_\mu = \left.\frac{p_\mu}{\text{max}(p_\mu)}\right|_{c_Z}~,
\end{equation}
by which we mean that for every $c_Z$, the corresponding set of $p_\mu$'s gets normalized by the maximal value of $p_\mu$ in that set. 
We choose $p_I = 0.9$, even though both slightly smaller and larger values around $p_I \approx 0.9$ perform equally well. However, going below $p_I \lesssim 0.8$ or above $p_I \gtrsim 0.95$ comes with different challenges. In the former (for large errors), we lose the important property that the sum of $p_\mu$'s decreases as a function of weight, $(\sum_\mu p _\mu)_{w=1} > (\sum_\mu p _\mu)_{w=2} > \dots$. In the latter (small errors), the range of values of $p_\mu$ is so large that one would need to use a 64-bit floating-point representation to compute the reward with sufficient precision. Since both RL algorithms and GPUs are currently designed to work best with 32-bit precision, we decide to avoid this range of values for $p_I$ during training, but we will still evaluate the strategies found by the RL agent on different values of $p_I$. 

\begin{figure}[ht!]
    \centering
    \includegraphics[width=0.4\textwidth]{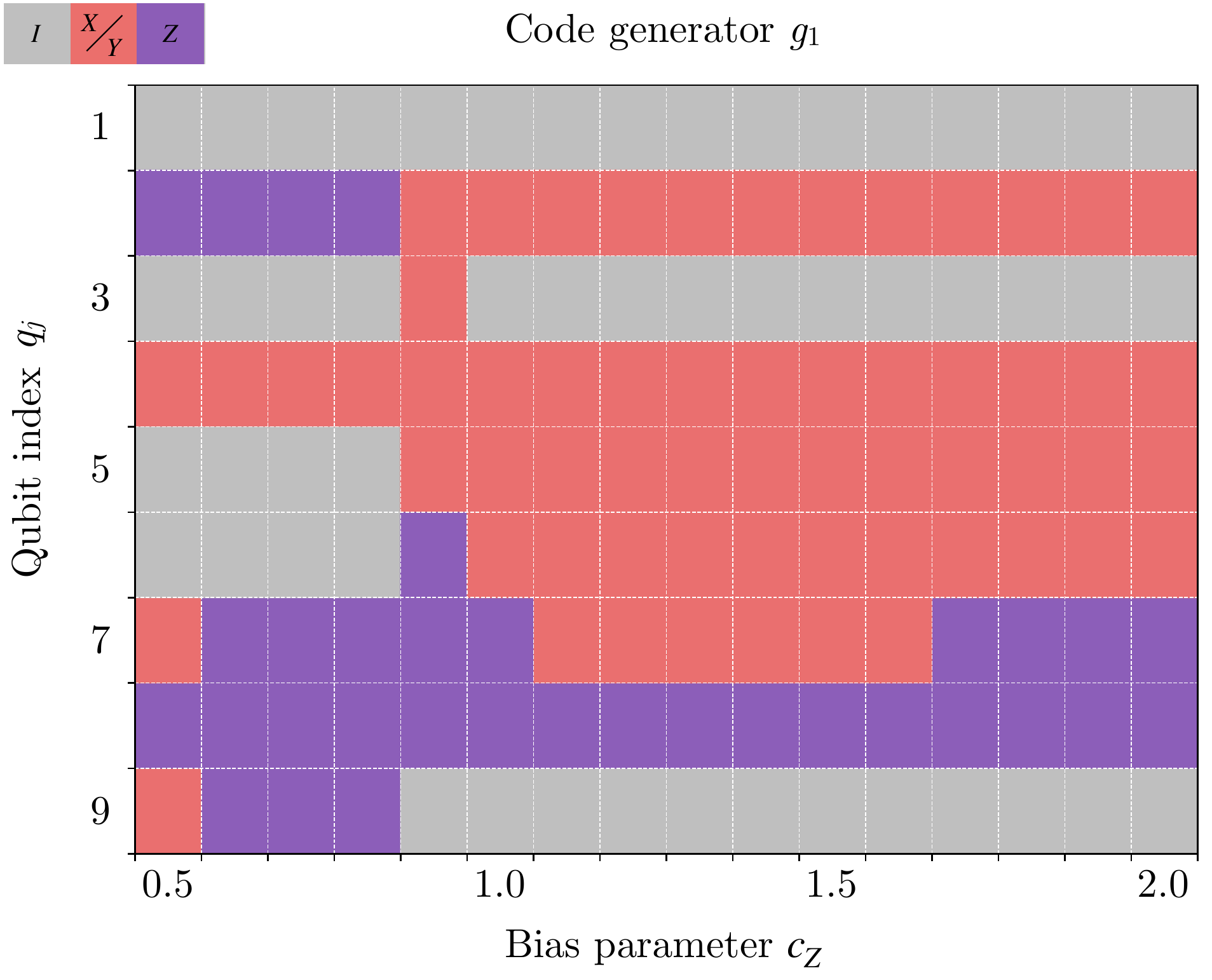}
    \caption{Evolution of the first code generator $g_1$ as a function of the bias parameter $c_Z$. This was evaluated at 16 discrete values of $c_Z$ even though the agent accepts continuous values of $c_Z$.}
    \label{fig:code_generator_1_vs_cz}
\end{figure}

We apply this strategy to target codes with parameters $n=9,~k=1$ in asymmetric noise channels. We allow a maximum number of 35 gates. Moreover, we consider an all-to-all connectivity, taking as available gate set $\{ H_i, S_i, \text{CNOT}(i,j)\}$, where $S_i$ is the phase gate acting on qubit $i$. 

As is the case for most RL learning procedures, every independent learning run will typically result in a different learned strategy by the agent. We will thus train many agents and post-select the few best performing ones. Now, there are in principle two different ways to make this selection: The first one is based on how well they minimize the weighted KL sum (which is what they were trained for). The second one is by evaluating the probability that a single error correction cycle will end in failure, i.e. the probability that the wrong correction would be applied based on the detected syndrome. More concretely, we classify the agents by summing these two quantities over $c_Z$. By evaluating them based on these two criteria we will be able to see whether agents trained with a computationally cheaper reward (the weighted KL sum) can be reused for the more complex task of minimizing the failure probability. To be precise, we define the failure probability as
\begin{equation}
    p_f = \sum_{\text{syn}} p( \text{correct wrong error}| \text{syn}) p(\text{syn})~, \label{eq:p_failure_definition}
\end{equation}
where syn is one of the possible $2^{n-k}$ syndromes, and where the error correction strategy is to always correct the most likely error to have happened given that specific syndrome. If more than one error is equally likely, we randomly select one of them.

\begin{figure*}[ht!]
    \centering
    \includegraphics[width=0.9\textwidth]{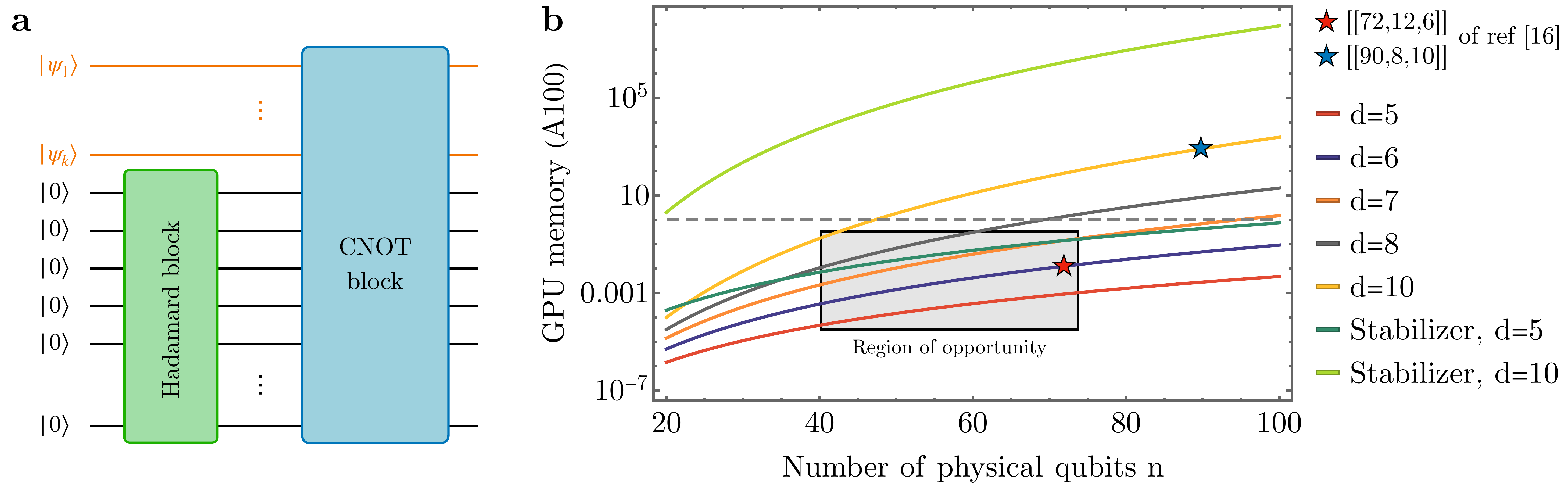}
    \caption{CSS code and encoding discovery. \textbf{a} Circuit architecture. Placing an initial block where only Hadamard gates are allowed, after which only CNOTs can be used, guarantees that the resulting code is CSS. In this work, we manually choose the content of the Hadamard block and ask the RL agent to find a suitable CNOT block. \textbf{b} Scaling to larger code parameters. We show the fraction of the 80 GB of GPU memory needed (NVIDIA A100 GPU) to store all the error operators that are required to reward the agent. We also show for comparison the memory load of stabilizer (non-CSS) code discovery for code distances $d=5$ and $d=10$. We identify a \textit{region of opportunity} where our RL strategy could outperform some of the qLDPC codes found in \cite{bravyi2023highthreshold} in the near future.}
    \label{fig:css_figure}
\end{figure*}

We discover codes with the following parameters: $\code{9,1,d_e(c_Z=0.5)=2}$, $\code{9,1,d_e(c_Z=0.6)=3}$, $\code{9,1,d_e(c_Z=1.4)=4}$, $\code{9,1,d_e(c_Z=2)=5}$. To the best of our knowledge, the last two codes are new. Codes inbetween, $0.5 \leq c_Z < 0.6$, have $d_e = 2$, $0.6 \leq c_Z < 1.4$ have $d_e = 3$, and so on. Even though all encodings that the agent outputs have circuit size 35, we notice that trivial gate sequences are applied at the last few steps, effectively reducing the overall gate count. We remark that this feature is not problematic: it means that the agent is done well before a new training run is launched, and the best thing it can do is collecting small negative rewards until the end. We manually prune the encodings to get rid of such trivial operations, and the resulting circuit sizes vary from 22 to 35, depending on the value of $c_Z$.

The main results are shown in Fig.~\ref{fig:noise_aware_result_9}. We start by comparing the two best-performing post-selected agents according to minimizing the weighted KL sum (green) and minimizing the failure probability (orange), see Fig.~\ref{fig:noise_aware_result_9}a,b. There we see that it is in general not true that minimizing the weighted KL sum will automatically correspond to a lower failure probability. Nevertheless, there is a nice correlation between the two, especially in the region $c_Z < 1$. More explicitly, the two best-performing agents show an almost identical performance in terms of both the weighted KL sum and the failure probability in this region. The situation changes when $c_Z \geq 1$, where we see that the optimal agents for the weighted KL sum are not optimal in terms of minimizing the failure probability, especially when $c_Z \geq 1.5$.

The results from Fig.~\ref{fig:noise_aware_result_9}a,b suggest that there are two branches separated by a kink at $c_Z=1$. Indeed, the two-branch structure can be easily understood in terms of the degree of asymmetry in the noise channel. For $c_Z < 1$, $Z-$errors are more likely than $X/Y-$errors, while the opposite happens for $c_Z > 1$. For $c_Z < 1$, a single error species had to be prioritized, but this number grows to two for $c_Z > 1$. At the middle point, $c_Z=1$, there is no preference for any of the three species. 

The last comparison between these two agents that we do is based on the smallest undetected effective weight of the codes found in Fig.~\ref{fig:noise_aware_result_9}c. The same behavior as before is observed: almost-perfect correlation up to $c_Z = 1$, and discrepancy at larger values of $c_Z$. Surprisingly, the code found by the best agent according to the weighted KL sum (green) at $c_Z=2$ has $d_e=5$, while the best code at minimizing the failure probability (orange) has $d_e = 4$. However, at the specific point $c_Z=2$ these two codes perform equally in terms of the failure probability (see Fig.~\ref{fig:noise_aware_result_9}b). An extended analysis of the performance of these two meta-agents, including a comparison against an ensemble of agents trained on a single value of $c_Z$ is included in Appendix~\ref{appendix:meta_agent_vs_ensemble}. 

Now we focus on the agent that performs best at minimizing the failure probability (orange) since it is the one of most interest in practical scenarios. We are specifically interested in doing a more detailed analysis to understand what are the characteristics of the codes that are being found. We begin by evaluating the performance of the \textit{same} agent on different values of $p_I$. This is shown in Fig.~\ref{fig:noise_aware_result_9}d, where we see that there is indeed a gain by encoding the quantum state, and it asymptotically follows a power law with exponent $\gtrsim 2$ depending on the specific value of $c_Z$.

We continue by analyzing the encoding circuits and code generators for some selected values of $c_Z$. These are chosen after computing the (symmetric) quantum weight enumerators according to Eq.~\eqref{eq:quantum-weight-enumerators}, which we show in Fig.~\ref{fig:noise_aware_circuits}c. There we see that the same code family is kept for $0.5 \leq c_Z < 0.9$, where Z errors are more likely than $X/Y$. From that point onward, the agent switches to a new code family that is kept until the end, $c_Z=2$. We thus choose to analyze the encoding circuits and their associated code generators for the values $c_Z = \{0.5, 0.9, 1.4,2\}$. 

We begin by showing the encoding circuits in Fig.~\ref{fig:noise_aware_circuits}a, highlighting common motifs that are re-used across various values of $c_Z$ with different colors, indicative of transfer learning. Another interesting behavior is that S gates are used more prominently at small values of $c_Z$, in particular in the combination $S\cdot H$. What this combination of gates does is a permutation: $X\to Y$, $Y \to Z$, $Z \to X$ (ignoring signs), which is very useful to exchange Y by Z efficiently. Moreover, we recall that the circuits that we show are pruned, i.e. trivial sequences of gates are manually eliminated. The agent actually uses lots of S gates at the very end of the encoding for almost all values of $c_Z$. This is no accident: we have constructed a scenario in which $X/Y$ errors are essentially indistinguishable, in the sense that they occur with equal probability. Thus the S gate (which exchanges X by Y) acts like an idling operation. This is cleverly leveraged by the agent to shorten the encoding circuits: once it cannot improve the final reward further, the strategy then becomes to reach that point as soon as possible and collect that final reward for as many steps as possible.

The final aspect we investigate is the code generators of such encoding circuits, which are shown in Fig.~\ref{fig:noise_aware_circuits}b. To aid visualization, we have chosen different colors for different Pauli matrices. However, since our scenario is by construction symmetric in $X/Y$, we choose to represent X and Y by the same color. Since the code used at $c_Z = 0.5$ is the only one from a different code family, it is natural that its code generator pattern is the most distinct. However, we see that the generators of the remaining values of $c_Z$ have similar structures at certain points. We also include in Fig.~\ref{fig:code_generator_1_vs_cz} how the first code generator $g_1$ (the first column in Fig.~\ref{fig:noise_aware_circuits}) changes across all values of $c_Z$. For instance, the transition between different code families is clearly visible at $c_Z=0.9$. In addition, other (not shown) features such as the agent using the \textit{same} circuit throughout $1.2 \leq c_Z \leq 1.6$ can also be glimpsed from Fig.~\ref{fig:code_generator_1_vs_cz} (even though one would need to analyze how all the generators evolve).

\section{Towards large code discovery}
\label{sec:towards}

It is estimated~\cite{Gidney_2021} that practical applications of quantum computers such as executing Shor's algorithm to factor 2048 RSA integers would require code distances in their 20s for physical gate error rates of order $10^{-3}$. Therefore, it is important to explore to which extent our RL-based strategy can be scaled up. In this Section, we will see that by restricting from stabilizer codes to CSS codes we are able to reduce the computational demands of our algorithms, leading to an estimated better scaling with larger code parameters.

A particularly useful subclass of stabilizer codes are \textit{Calderbank-Shor-Steane (CSS) codes}~\cite{Steane1996,Shor1996}. They are defined by their stabilizer generators containing either only $X$ or only $Z$ Pauli operators. This restriction is useful because $X$-type and $Z$-type errors are detected \textit{independently}, thereby \textit{implying} the detection of Y-type errors when the corresponding X and Z-type stabilizers fire simultaneously. Moreover, strong contenders for implementation in large-scale quantum computations such as surface codes or color codes are of the CSS type. 

The independence of X and Z-type error detection means that the number of error operators that we have to keep track of drastically reduces from \eqref{eq:numE} to
\begin{equation}
    |\left\{E_\mu\right\}^{\text{CSS}}|_{w \leq d-1} = 2 \sum_{w=0}^{d-1} \binom{n}{w}~, \label{eq:numE_CSS}
\end{equation}
where the overall factor of 2 counts both X and Z-type errors. Crucially, there is roughly an exponential factor in $d$ less number of errors than in the case of general stabilizer codes \eqref{eq:numE} that have to be considered.

Motivated by this reduction, we present a strategy that will exclusively target CSS code discovery. In order to implement such a restriction while being able to codiscover the encoding circuit simultaneously, it is sufficient to constrain the structure of the circuit. Indeed, if the circuit is built from an initial block of Hadamard gates applied to a subset of the qubits followed by CNOT gates thereafter guarantees that the resulting code will be CSS (see Appendix~\ref{appendix:CSS} for a proof and Fig.~\ref{fig:css_figure}a for a schematic representation). This can be seen intuitively by the fact that a Hadamard gate exchanges an $X$ by a $Z$ (and viceversa), and CNOT gates propagate those Paulis to different positions within that stabilizer generator (without mixing $X$'s and $Z$'s).

There are several possible modifications that we could do to our RL strategy, even though here we will only contemplate two. The first one would be to keep as actions both H and CNOT gates for the agent to use, but penalize the agent every time that a Hadamard gate is used after a CNOT gate. This would in principle lead to an agent that would know what is the correct architecture to be used for CSS codes at expenses of having to fine-tune this new penalty term in the reward. The second option would be to only allow the agent to use CNOT gates and \textit{we} would be the ones deciding where to place the initial Hadamard gates. The advantage of this strategy with respect to the first one is that the search space of strategies for the agent is greatly reduced and the learning procedure would be much easier. 

Since the purpose of this Section is to explore the potential of an idea rather than giving definitive results, we choose the mixed \textit{human-AI} strategy where \textit{we} are the ones deciding the content of the Hadamard block and where the agent has to discover suitable CNOT blocks. In this way, we simplify the task of the agent as much as possible. For a schematic picture of this approach we refer the reader to Fig.~\ref{fig:css_figure}a. 

Next, we make some estimations on the practical limits of CSS code discovery using a KL-based reward. As we have seen, a crucial ingredient of efficient QEC code discovery driven by RL is being able to both simulate the environment and train the RL agent with GPUs. With this in mind, we estimate the amount of memory that would be needed to store all error operators for some code parameters $n$ and $d$ (this calculation is independent of $k$). This estimation amounts to counting the number of error operators \eqref{eq:numE_CSS}, times the amount of binary digits that have to be specified for each of them. We show the results of this estimation in Fig.~\ref{fig:css_figure}b for code distances from 5 to 10 and physical qubit numbers of 20 to 100. In particular, we consider what fraction of memory they would occupy in an NVIDIA A100 GPU, which is the modern GPU model standard (see Appendix~\ref{appendix:gpu_memory_estimation} for all the details of our estimation). The results shown in Fig.~\ref{fig:css_figure}b indicate that our approach can be extended to $\sim 100$ physical qubit numbers ($d=6$) and to approximately 40 physical qubits and $d=10$ in a \textit{single} GPU. Moreover, we identify a \textit{region of opportunity} that could potentially lead to new codes surpassing the properties of the smaller qLDPC codes found in \cite{bravyi2023highthreshold} since we do not have an ansatz that limits the families of codes that we could find. We emphasize that not only would we discover the code, but the encoding circuit would also be simultaneously discovered.

We have tested this approach by targeting weakly self-dual codes (meaning $\text{num}(H) = (n-k)/2$) of distance $d=5$ using a next-to-nearest neighbor CNOT connectivity (with periodic boundary conditions) and initial Hadamard gates in alternating qubit indices. We have found that we can discover $\code{17,1,5}$ codes (with $\text{num}(H)=8$) in roughly 20 minutes, \textit{from scratch} and with their encoding circuit. An example of such a discovered circuit is the following:

\scalebox{1.0}{
\Qcircuit @C=0.2em @R=0.1em @!R { \\
	 	\nghost{{q}_{0} :  } & \lstick{{q}_{0} :  } & \qw & \qw & \qw & \targ & \ctrl{15} & \qw & \targ & \qw & \qw & \qw & \ctrl{15} & \qw & \qw & \qw & \qw & \qw & \qw & \qw & \qw\\
	 	\nghost{{q}_{1} :  } & \lstick{{q}_{1} :  } & \qw & \qw & \targ & \qw & \qw & \targ & \ctrl{-1} & \qw & \targ & \qw & \qw & \qw & \ctrl{15} & \qw & \qw & \qw & \qw & \qw & \qw\\
	 	\nghost{{q}_{2} :  } & \lstick{{q}_{2} :  } & \gate{\mathrm{H}} & \ctrl{1} & \qw & \ctrl{-2} & \qw & \qw & \qw & \qw & \qw & \qw & \qw & \qw & \qw & \qw & \qw & \qw & \qw & \qw & \qw\\
	 	\nghost{{q}_{3} :  } & \lstick{{q}_{3} :  } & \qw & \targ & \qw & \targ & \qw & \ctrl{-2} & \qw & \targ & \ctrl{-2} & \targ & \qw & \qw & \qw & \qw & \qw & \ctrl{2} & \qw & \qw & \qw\\
	 	\nghost{{q}_{4} :  } & \lstick{{q}_{4} :  } & \gate{\mathrm{H}} & \ctrl{1} & \qw & \ctrl{-1} & \qw & \qw & \targ & \qw & \qw & \qw & \qw & \qw & \qw & \qw & \qw & \qw & \qw & \qw & \qw\\
	 	\nghost{{q}_{5} :  } & \lstick{{q}_{5} :  } & \qw & \targ & \qw & \targ & \qw & \qw & \qw & \ctrl{-2} & \targ & \ctrl{-2} & \qw & \qw & \qw & \qw & \targ & \targ & \qw & \qw & \qw\\
	 	\nghost{{q}_{6} :  } & \lstick{{q}_{6} :  } & \gate{\mathrm{H}} & \ctrl{1} & \qw & \ctrl{-1} & \qw & \targ & \ctrl{-2} & \qw & \qw & \targ & \qw & \qw & \qw & \ctrl{1} & \qw & \qw & \qw & \qw & \qw\\
	 	\nghost{{q}_{7} :  } & \lstick{{q}_{7} :  } & \qw & \targ & \qw & \qw & \qw & \qw & \targ & \qw & \ctrl{-2} & \qw & \qw & \ctrl{2} & \qw & \targ & \ctrl{-2} & \qw & \qw & \qw & \qw\\
	 	\nghost{{q}_{8} :  } & \lstick{{q}_{8} :  } & \gate{\mathrm{H}} & \ctrl{1} & \qw & \qw & \qw & \ctrl{-2} & \qw & \targ & \qw & \ctrl{-2} & \qw & \qw & \qw & \qw & \qw & \ctrl{2} & \targ & \qw & \qw\\
	 	\nghost{{q}_{9} :  } & \lstick{{q}_{9} :  } & \qw & \targ & \qw & \targ & \qw & \qw & \ctrl{-2} & \qw & \targ & \qw & \qw & \targ & \qw & \qw & \qw & \qw & \ctrl{-1} & \qw & \qw\\
	 	\nghost{{q}_{10} :  } & \lstick{{q}_{10} :  } & \gate{\mathrm{H}} & \ctrl{1} & \qw & \ctrl{-1} & \qw & \qw & \targ & \ctrl{-2} & \qw & \ctrl{2} & \qw & \qw & \qw & \targ & \targ & \targ & \targ & \qw & \qw\\
	 	\nghost{{q}_{11} :  } & \lstick{{q}_{11} :  } & \qw & \targ & \qw & \qw & \qw & \targ & \qw & \qw & \ctrl{-2} & \qw & \qw & \targ & \qw & \ctrl{-1} & \qw & \qw & \ctrl{-1} & \qw & \qw\\
	 	\nghost{{q}_{12} :  } & \lstick{{q}_{12} :  } & \gate{\mathrm{H}} & \ctrl{1} & \qw & \qw & \qw & \qw & \ctrl{-2} & \targ & \qw & \targ & \qw & \qw & \qw & \qw & \ctrl{-2} & \qw & \qw & \qw & \qw\\
	 	\nghost{{q}_{13} :  } & \lstick{{q}_{13} :  } & \qw & \targ & \qw & \targ & \qw & \ctrl{-2} & \qw & \qw & \targ & \ctrl{1} & \qw & \ctrl{-2} & \qw & \qw & \qw & \qw & \qw & \qw & \qw\\
	 	\nghost{{q}_{14} :  } & \lstick{{q}_{14} :  } & \gate{\mathrm{H}} & \ctrl{1} & \qw & \ctrl{-1} & \qw & \targ & \qw & \ctrl{-2} & \qw & \targ & \qw & \qw & \qw & \qw & \qw & \qw & \qw & \qw & \qw\\
	 	\nghost{{q}_{15} :  } & \lstick{{q}_{15} :  } & \qw & \targ & \qw & \qw & \targ & \qw & \ctrl{1} & \qw & \ctrl{-2} & \qw & \targ & \qw & \qw & \qw & \qw & \qw & \qw & \qw & \qw\\
	 	\nghost{{q}_{16} :  } & \lstick{{q}_{16} :  } & \gate{\mathrm{H}} & \qw & \ctrl{-15} & \qw & \qw & \ctrl{-2} & \targ & \qw & \qw & \qw & \qw & \qw & \targ & \qw & \qw & \qw & \qw & \qw & \qw\\
\\ }}

It consists of 8 Hadamard gates (that we chose) and a remaining sequence of 46 CNOT gates discovered by the agent. An interesting  strategy that the agent uses is first building Bell pairs between adjacent qubits (which are $\code{2,0,2}$ codes) and then entangle these pairs with each other to gradually build up a $d=5$ code.

We remind the reader that the largest (non-CSS) code that we had shown in previous sections was $\code{15,2,5}$ and it needs roughly 4 hours of computing. Encouraged by this result, we have tried targeting $n \geq 20$ and $d \geq 6$. The result is that while codes with $d=5$ have been found, we have been unsuccessful with $d\geq6$. Our training curves suggest that the loss landscape of this problem is quite complex and an auxiliary, conceptually novel strategy that allows us to navigate such a vast space of circuits in a more effective way is needed. In any case, we leave this study for future work.

\section{Conclusions and Outlook} \label{sec:conclusion}

We have presented an efficient RL framework that is able to simultaneously discover QEC codes and their encoding circuits from scratch, given a qubit connectivity, gate set, and error operators. It learns strategies simultaneously for a range of noise models, thus re-using and transferring discoveries between different noise regimes. We have been able to discover codes and circuits up to 20 physical qubits and code distance 5. This is thanks to our formulation in terms of stabilizers, that serve both as compact input to the agent as well as the basis for rapid Clifford simulations, which we implemented in a vectorized fashion using a modern machine-learning framework.

In the present work, we have focused on the quantum communication or quantum memory scenario, where the encoding circuit itself can be assumed error-free since we focus on errors happening during transmission. As a result, our encoding circuits are not fault tolerant, i.e. single errors, when introduced, might sometimes proliferate to become incorrigible. Flag-based fault tolerance \cite{Chao_2018} added on top of our encoding circuits could turn them fault tolerant. 

We have also shown how to efficiently scale up this strategy by exclusively targeting CSS codes, potentially being able to outperform the recent \textit{quasi-cyclic} codes from~\cite{bravyi2023highthreshold} in the near future. In addition, not only would the codes be discovered, but their encoding circuits would also be automatically known.

One of the limits of our approach is GPU memory. However, this could be circumvented through different means. While it is always possible to trade performance by memory load, the tendency to train very large AI models is thrusting both the development of novel hardware with increased memory capabilities and the integration of distributed computing options in modern machine learning libraries. These developments makes us envision scenarios where the framework presented in this work could be scaled up straightforwardly. This makes us optimistic about the future of AI-discovered QEC in the very near future.

\begin{acknowledgments}
Fruitful discussions with Sangkha Borah, Jonas Landgraf, Maximilian Naegele and Oleg Yevtushenko are thankfully acknowledged. We are thankful to Markus Grassl for comments on the first version of this manuscript.
This research is part of the Munich Quantum Valley, which is supported by the Bavarian state government with funds from the Hightech Agenda Bayern Plus.
\end{acknowledgments}

\bibliography{main}

\appendix

\section{Elementary example: The three qubit repetition code.}
\label{appendix:three_qubit_repetition_code_example}

\begin{figure}[ht!]
    \centering
    \includegraphics[width=0.4\textwidth]{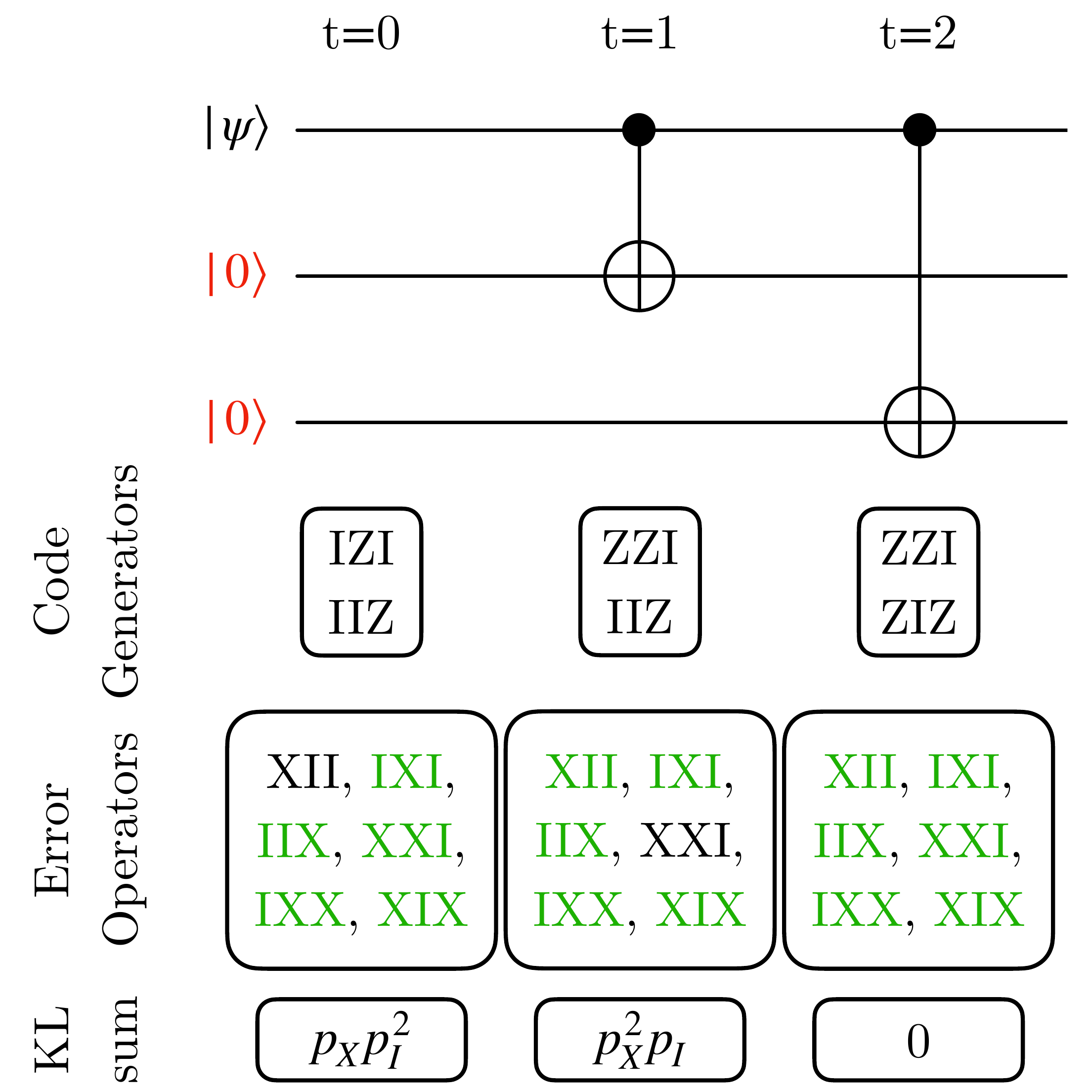}
    \caption{Encoding process of the three-qubit repetition code. The first qubit can be in an arbitrary state $|\psi\rangle$ and it will be transformed into the logical state $|\psi\rangle_L$ once the encoding has finished. We show how the code generators change as gates are applied, and which of the error operators can be detected (green) using the code generators available at every timestep. The KL sum quantifies how close the KL conditions are of being satisfied and is defined in \eqref{eq:KL_sum_def}. The encoding terminates successfully if the KL sum is zero.}
    \label{fig:3_qubit_example}
\end{figure}

Here we illustrate the encoding process of the three qubit repetition code, which is a $\code{3,1}$ code that can correct single-qubit bit flips. The whole procedure is summarized in Fig.~\ref{fig:3_qubit_example}.

There are two code generators and the targeted error set consists of all weight$-1$ and weight$-2$ bit-flip (Pauli$-X$) operators: $XII, IXI, IIX, XXI, XIX, IXX$. We remind the reader that detecting this set implies being able to correct all single-qubit bit-flips. The generators start being $IZI, IIZ$ in the absence of gates, and transform into $ZZI, ZIZ$ after applying two CNOT gates with control on the first qubit and target on the second and third qubits (see Fig.~\ref{fig:3_qubit_example}). Initially, only the error operator $XII$ is not detected (the other error operators can be checked to satisfy the Knill-Laflamme (KL) conditions \eqref{eq:KL_anticommute},~\eqref{eq:KL_in_S}). After the first CNOT gate, the undetected error becomes $XXI$, and after the second CNOT all target errors are detected. 

In order to have a single number that quantifies to which degree the KL conditions are not satisfied, we introduce a quantity that we call the KL sum,
\begin{equation}
    \sum_\mu p_\mu K_\mu~, \label{eq:KL_sum_def}
\end{equation}
with the index $\mu$ running over the number of operators and where $K_\mu$ is 0 if the corresponding error $E_\mu$ is detected and 1 otherwise, with $p_\mu$ being the occurrence probability of $E_\mu$. The KL sum is closely related to the reward that we have used in our RL framework (see \eqref{eq:reward}) and is a useful indicator of how close a gate sequence is to be able to detect all the errors that were chosen. 

In the 3-qubit repetition code under consideration (see Fig.~\ref{fig:3_qubit_example}), the KL sum starts being $p_X p_I^2$ because only the error operator $XII$ is not detected, changing to $p_X^2 p_I < p_X p_I^2$ after application of the first CNOT gate. When all KL conditions are satisfied, the KL sum is zero and all error operators that were considered can be detected, leading to a successful encoding.

\section{Implementation details of the vectorized Clifford simulator}
\label{appendix:simulations}

\begin{figure}
    \centering
    \includegraphics[width=0.4\textwidth]{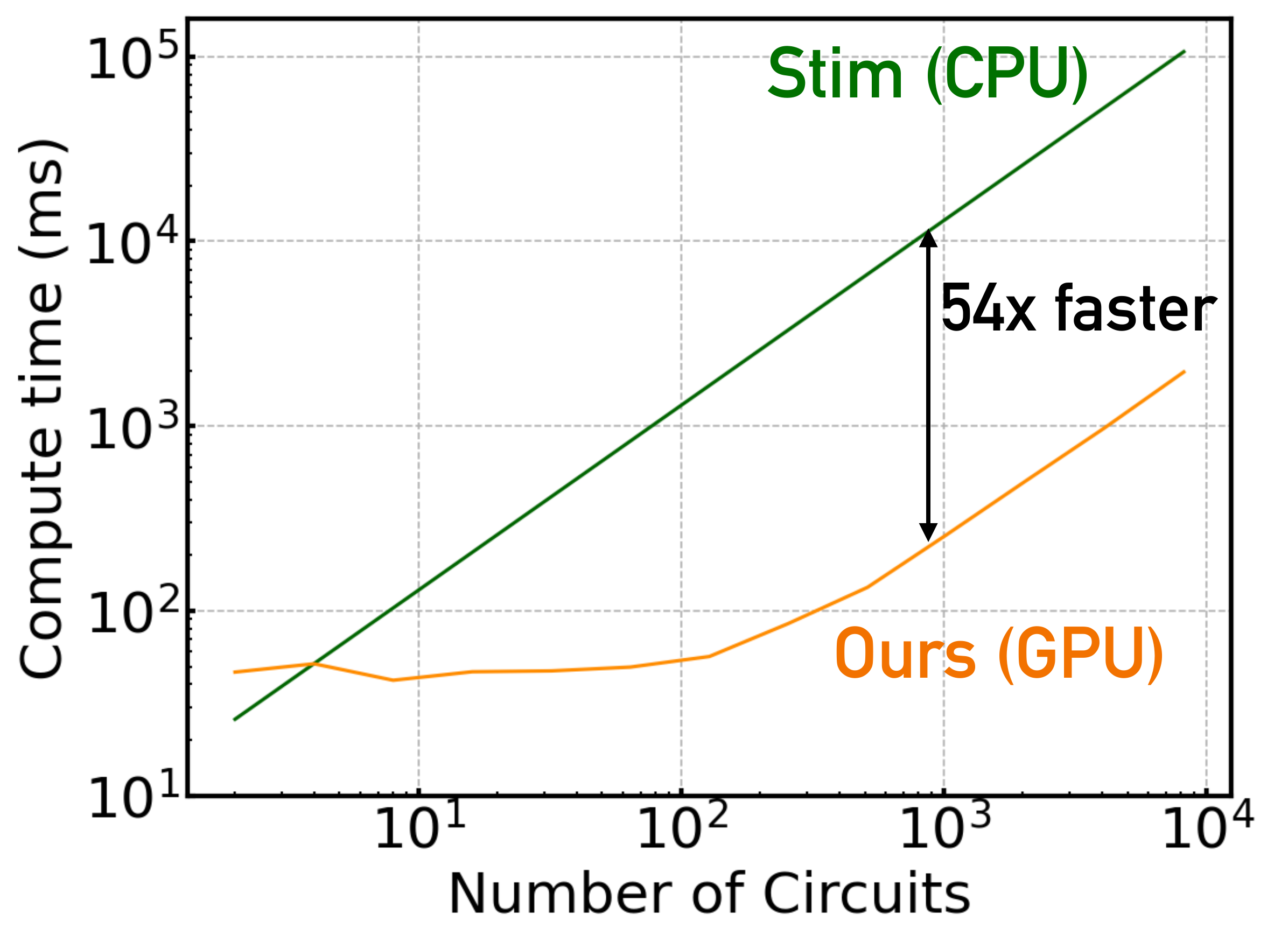}
    \caption{Compute time required to generate random Clifford circuits of 1000 gates for 40 qubits with \textsc{Stim} (green) and with our simulator (orange).}
    \label{fig:stim_vs_qdx}
\end{figure}

Here we give more details on the implementation of our simulations, which are based on the binary symplectic formalism of the Pauli group and that have been optimized to be compatible with modern vectorized machine learning frameworks running on Graphical Processing Units (GPU). In particular, we ignore global phases in our approach, since they play no role in the KL conditions Eq.~\eqref{eq:KL_anticommute},~\eqref{eq:KL_in_S}.

A stabilizer generator $g_i$ is formally represented as a Pauli string $P_1 \otimes P_2 \otimes \dots \otimes P_n$, where $P_i \in \{I,X,Y,Z\}$ is any Pauli operator, and numerically as a binary vector of size $2n$. For example, the Pauli matrices are represented as $I = (0, 0)$, $X = (1, 0)$, $Y = (1,1)$, $Z = (0,1)$, and a general Pauli string is represented as $(x_{1}, \dots,  x_{n}, \; z_{1}, \dots, z_{n})$, where all $x_i$ and $z_i$ are either 0 or 1. For instance, the binary vector $(1,1,0,0,\;0,1,1,0)$ represents the Pauli string $XYZI$. Matrix multiplication gets mapped to binary sum (ignoring global phases), e.g.
\begin{equation}
    X \cdot Y = Z \longleftrightarrow (1,0) + (1,1) = (0,1)~(\text{mod}2)~.
\end{equation}

A stabilizer code is specified by $n-k$ stabilizer group generators $S_\mathcal{C} = \langle g_1, g_2, \dots , g_{n-k} \rangle$ and is therefore represented by a \textit{check matrix} $G$~\cite{Aaronson_2004}, which is a $(n-k) \times 2n$ binary matrix where each row $i$ represents the Pauli string $g_i$ from $S_\mathcal{C}$. Clifford gates map Pauli strings to Pauli strings, meaning that a check matrix $G$ gets mapped to a different check matrix $G'$ under the action of any Clifford gate. It is sufficient to consider the action of the Clifford gates H,S,CNOT on $X/Z$ stabilizers. For instance, the action of H is the well-known
\begin{equation}
    H  X  H = Z,\; H  Z  H = X~,
\end{equation}
meaning that it exchanges $X$ by $Z$. More generally, $H_i$ exchanges columns $i$ and $i+n$ of a check matrix $G$. We implement this transformation by representing $H_i$ with a binary matrix $H(i)_b$ and by performing binary matrix multiplication between $G$ and $H(i)_b$. Explicitly, $H(i)_b$ is the $2n \times 2n$ identity matrix with columns $i$ and $i+n$ exchanged,
\begin{equation}
H(i)_b = 
    \begin{pmatrix}
    1 & 0 & \cdots & 0 & \cdots & 0 & \cdots & 0 \\
    0 & 1 & \cdots & 0 & \cdots & 0 & \cdots & 0 \\
    0 & 0 & \cdots & 0 & \cdots & 0 & \cdots & 0 \\
    \vdots & \vdots &  & \vdots &  & \vdots &  & \vdots \\
    0 & 0 & \cdots & 0 & \cdots & \underbrace{1}_{i} & \cdots & 0 \\
    \vdots & \vdots &  & \vdots &  & \vdots &  & \vdots \\
    0 & 0 & \cdots & \underbrace{1}_{i+n} & \cdots & 0 & \cdots & 0 \\
    \vdots & \vdots &  & \vdots &  & \vdots &  & \vdots \\
    0 & 0 & \cdots & 0 & \cdots & 0 & \cdots & 1 \\
\end{pmatrix}~,
\end{equation}
and matrix multiplication must be done from the right, i.e. $G' = G \cdot H(i)_b~(\text{mod}2)$. Binary matrix representations can be built for all $S_i$ and $\text{CNOT}(i,j)$ gates in a similar manner.

Two Pauli strings $P_1$ and $P_2$ either commute or anticommute. We compute this by evaluating the binary symplectic bilinear
\[
 P_1 \cdot \Omega \cdot P_2^T = \left\{
  \begin{array}{ll}
    0 & \text{if } P_1 \text{ and } P_2 \text{ commute}, \\
    1 & \text{if } P_1 \text{ and } P_2 \text{ anticommute}
  \end{array}
\right.
\]
where $P_1$ and $P_2$ are the corresponding binary representations and $\Omega$ is the $2n \times 2n$ symplectic metric
\begin{equation}
    \Omega = 
    \begin{pmatrix}
    \mathbb{0}_n & \mathbb{1}_n \\
    \mathbb{1}_n & \mathbb{0}_n \\
    \end{pmatrix}~.
\end{equation}
In our problem, we want to determine whether a list of operators $\{E_\mu\}$ anticommute with any of the code generators $g_i$. We group the error operators inside a binary matrix $E_M$, where each row corresponds to the binary representation of a different operator, and we compute
\begin{equation}
    E_M \cdot \Omega \cdot G^T~.
\end{equation}
The result is a binary matrix with dimensions $(\text{num}(E_\mu), n-k)$. The first KL condition Eq.~\eqref{eq:KL_anticommute} requires checking whether \textit{at least} one code generator $g_i$ anticommutes with any given error operator. This means that the result has to be transformed into a binary array of dimension $\text{num}(E_\mu)$, where a 1 means that the first KL condition Eq.~\eqref{eq:KL_anticommute} is satisfied for the corresponding operator $E_\mu$ and that is zero otherwise.

The second KL condition Eq.~\eqref{eq:KL_in_S} requires checking whether any error operator $E_\mu \in S_\mathcal{C}$. In principle, the full stabilizer group of $2^{n-k}$ elements must be built at every time step of our simulations. For the physical qubit numbers that we have considered in our work, this computation is still fast enough, becoming more challenging as $n-k \geq 13$. In practice, not many error operators end up being in $S_\mathcal{C}$, which we leverage by introducing a \textit{softness} parameter $s$ such that only a subgroup of $S_\mathcal{C}$ is built. More precisely, $s=0$ means that this subgroup is empty, $s=1$ means taking only the generators $g_i$ as the subgroup, $s=2$ means taking the generators $g_i$ and all pairwise combinations of generators $g_i g_j$, and so on for larger $s$.

In the end, all the operations that are required for both simulating the quantum circuits and to compute the reward have been implemented using binary linear algebra (matrix-vector multiplication). These are efficiently vectorizable operations using modern machine learning frameworks such as \textsc{Jax}~\cite{jax2018github} and are extremely fast on a GPU. In particular, we compare the compute time needed to evolve random Clifford circuits of 1000 gates on 40 qubits with our simulator and with \textsc{Stim}~\cite{gidney2021stim} in Fig.~\ref{fig:stim_vs_qdx}, and we see that in this regime we are around 50 times faster than \textsc{Stim}.

\section{RL algorithm hyperparameters and performance}
\label{appendix:hyperparameters}

\begin{figure}
    \centering
    \includegraphics[width=0.4\textwidth]{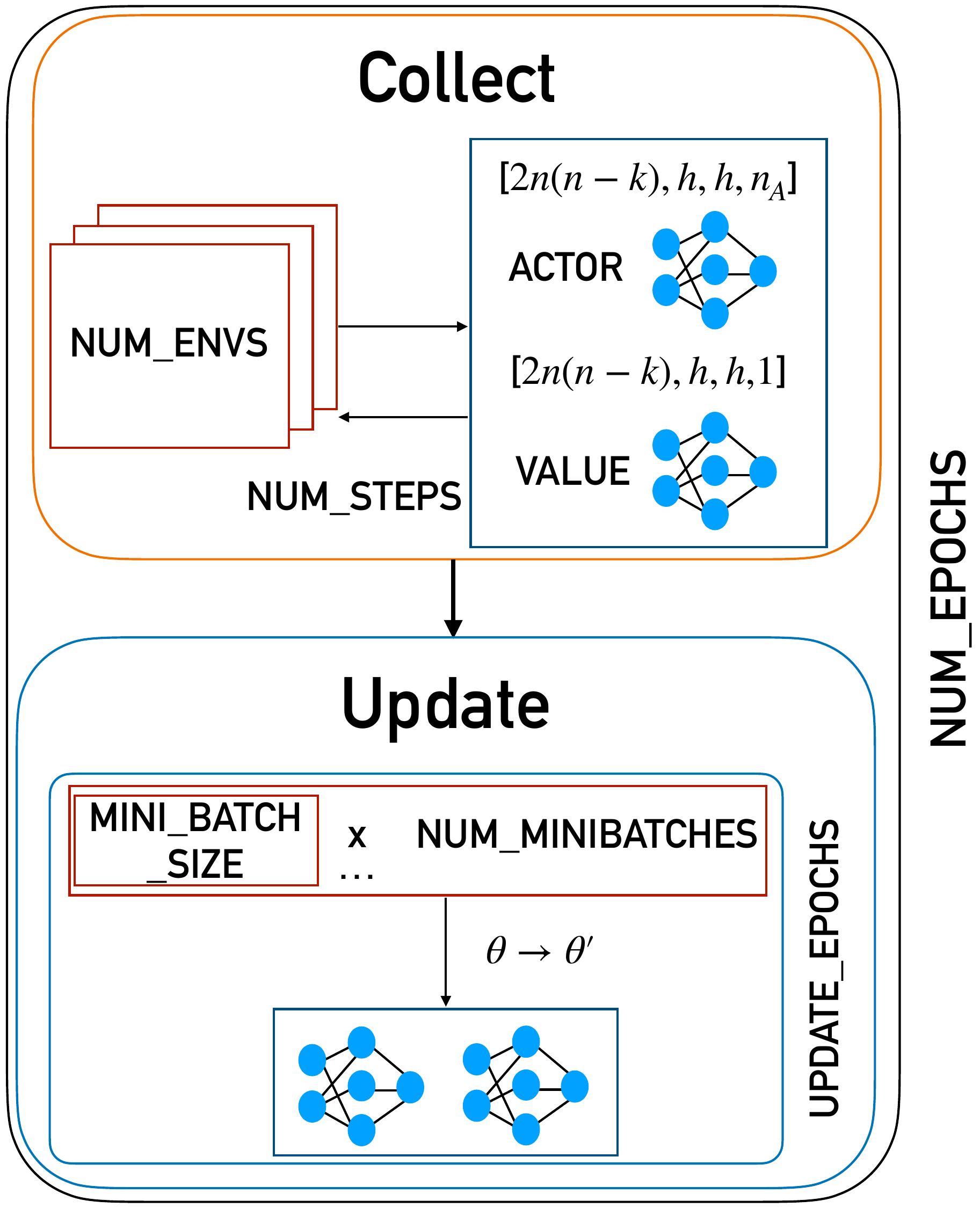}
    \caption{Configuration of the PPO algorithm used in this work, focusing on its structural and operational aspects. In the Collect phase, the agent interacts with the environments to extract triples (observation, action, reward) that are then used in the Update phase to update the parameters of the neural networks via stochastic gradient descent.}
    \label{fig:ppo_algorithm}
\end{figure}

\begin{figure*}
    \centering
    \includegraphics[width=0.9\textwidth]{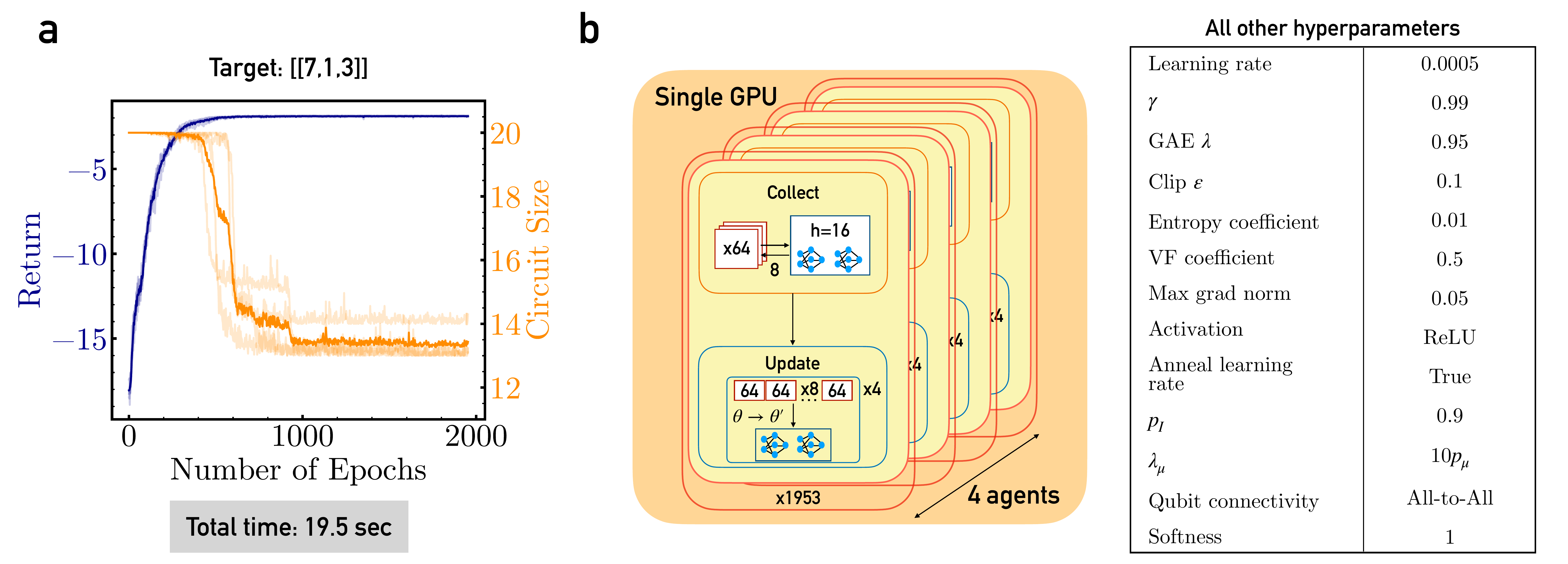}
    \caption{Example of a training trajectory for $\code{7,1,3}$ code discovery (a) with the complete set of hyperparameters that were chosen (b). Here, 4 parallel agents each interact with batches of 64 circuits processed in parallel. Each agent finds a different encoding circuit, and the training finishes in 20 sec on a single GPU. The meaning of every hyperparameter is explained in Appendix~\ref{appendix:hyperparameters}, while the softness parameter is explained in Appendix~\ref{appendix:simulations}.}
    \label{fig:training_example}
\end{figure*}

In this work we use a state-of-the-art policy-gradient method called Proximal Policy Optimization (PPO)~\cite{schulman2017proximal}. It is an actor-critic method with two separate entities that we implement as neural networks: the policy network predicts action probabilities (which quantum gate should be applied next) and the critic network predicts the advantage of that action (how valuable it was). 

We use the PPO implementation of \cite{lu2022discovered}, which we will break down in more detail (see also Fig.~\ref{fig:ppo_algorithm} and Table~\ref{table:hyperparameter_values} for a list of hyperparameters). In our implementation, the RL environment is vectorized, meaning that the agent interacts with multiple different quantum circuits at the same time. The hyperparameter that determines this number of RL environments is called $\text{NUM}\_\text{ENVS}$. The learning algorithm consists of two processes: collect and update. During collection, the agent interacts with the environments and a total of $\text{NUM}\_\text{STEPS}$ sequences of (observation, action, reward) are collected per environment. Following the collection, the update process begins. Here, we have a total of $\text{NUM}\_\text{ENVS} * \text{NUM}\_\text{STEPS}$ individual steps that are shuffled and reshaped into $\text{NUM}\_\text{MINIBATCHES}$ minibatches (each of size $\text{BATCH}\_\text{SIZE}$). These are used for updating the weights of the neural networks through stochastic gradient descent, which happens a number of $\text{UPDATE}\_\text{EPOCHS}$ times during every update process. The whole collection-update cycle gets repeated NUM\_EPOCHS times.

The neural networks that we have chosen are standard feedforward fully-connected neural networks with identical architectures for both the actor and value networks. In particular, they both consist of an input layer of size $2n(n-k)$ given by the observation from the environment, followed by two hidden layers of size $h$ and an output layer of size $n_A$ (number of actions) in the case of the actor network and of size 1 for the value network (see Fig.~\ref{fig:ppo_algorithm}). The number of actions $n_A$ is determined by the available gate set and qubit connectivity.

Other hyperparameters that participate in the PPO implementation which we include for completeness (but that we refer to \cite{schulman2017proximal} for further explanations) are the discount factor $\gamma$, the generalized advantage estimator (GAE) parameter $\lambda$, the actor loss clipping parameter $\varepsilon$, the entropy coefficient and the value function (VF) coefficient.

Regarding the optimizer itself, we use ADAM with a clipping in the norm of the gradient (MAX\_GRAD\_NORM) and some initial learning rate (LR) that gets annealed (ANNEAL\_LR) using a linear schedule as the training evolves, see Table~\ref{table:hyperparameter_values} for specific numerical values.

Finally, our actor and critic networks are chosen to be identical with two layers of between 16 and 400 neurons in each hidden layer and a ReLU activation function.
\begin{figure}
    \centering
    \includegraphics[width=0.4\textwidth]{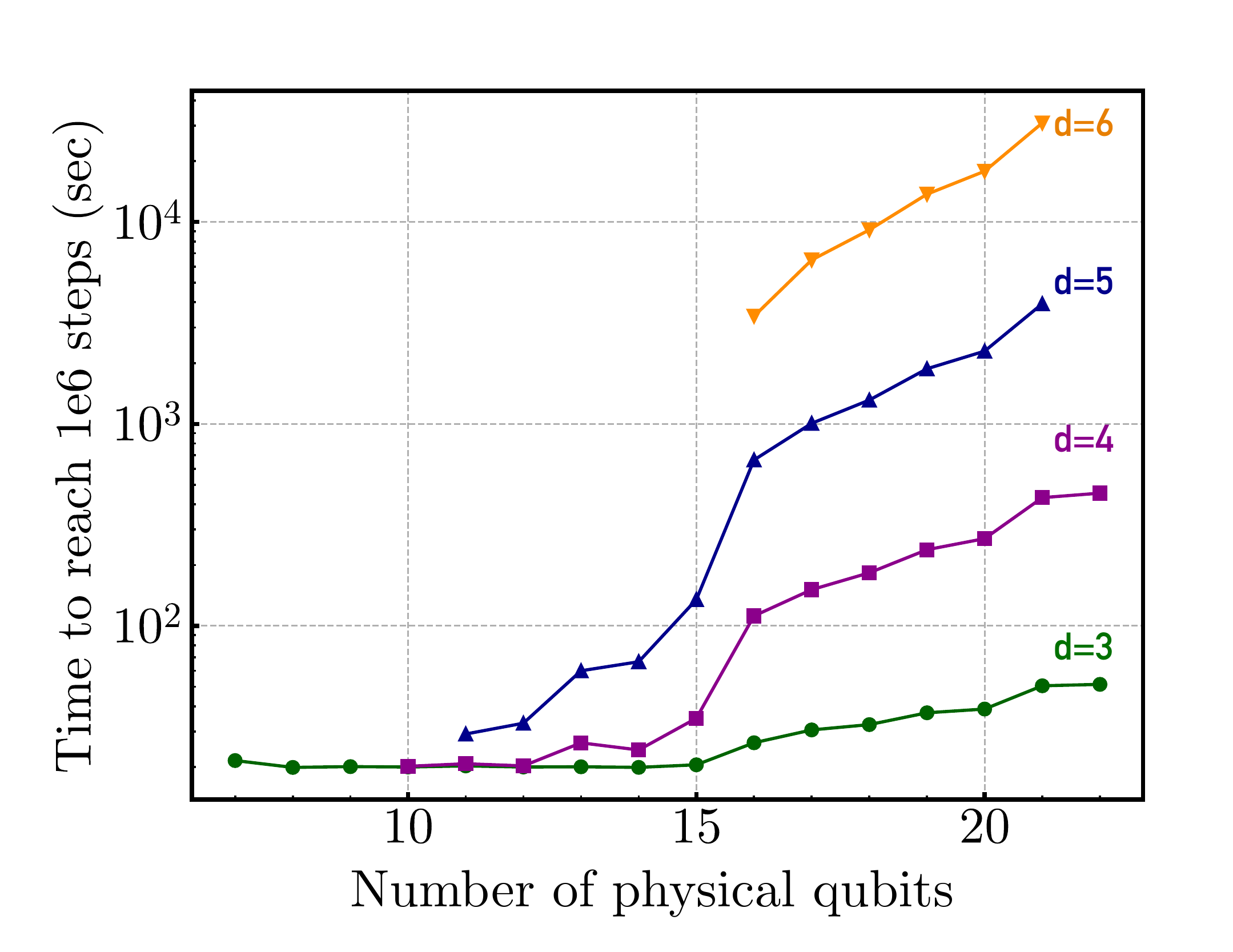}
    \caption{Execution time of training trajectories of 4 parallel agents (in a single GPU) with identical hyperparameters as those shown in Fig.~\ref{fig:training_example} with different number of physical qubits $n$ and code distance $d$ (but keeping the number of logical qubits $k=1$).}
    \label{fig:scaling_4_agents}
\end{figure}
Next, we show an example of a typical training trajectory in Fig.~\ref{fig:training_example} together with all the hyperparameter numerical values that were used and the execution time on a single NVIDIA Quadro RTX 6000 GPU. There, 4 agents are tasked to find \(\code{7,1,3}\) codes, which each of them completes successfully running in parallel in 20 sec. The error channel is chosen to be global symmetric depolarizing with $p_I=0.9$ (i.e. $p_X=p_Y=p_Z= 1-p_I/3$). The average circuit size starts being 20 by design, i.e. if no code has been found after 20 gates, the circuit gets reinitialized. This number starts decreasing when codes start being found and it saturates to a final value, which is in general different for each agent. As a final remark, running the same script on a CPU node with two Xeon Gold 6130 processors takes 7 min 40 sec. 

Finally, we show how the runtime scales when increasing the number of physical qubits $n$ and the code distance $d$. In order to get a meaningful comparison, we fix all other hyperparameters to be identical to those shown in Fig.~\ref{fig:training_example}. We remark that in general the agents will not have converged to a successful encoding sequence given the allotted resources.

\begin{table}[h!]
\centering
\begin{tabular}{l*{2}{ccc}}
\hline
\textbf{Hyperparameter}      & \textbf{Value}    \\ \hline
LR                      & $(1-5)\times 10^{-4}$     \\ 
NUM\_ENVS               & 8-1024               \\ 
NUM\_STEPS              & 8-32                \\ 
NUM\_EPOCHS        & 1000-12000               \\ 
UPDATE\_EPOCHS          & 2-4                 \\ 
NUM\_MINIBATCHES        & 8-128               \\ 
GAMMA                   & 0.99              \\ 
GAE\_LAMBDA             & 0.95              \\ 
CLIP\_EPS               & 0.1-0.2               \\ 
ENT\_COEF               & 0.01-0.05              \\ 
VF\_COEF                & 0.5               \\ 
MAX\_GRAD\_NORM         & 0.05-0.25               \\ 
ANNEAL\_LR              & True              \\ \hline
\end{tabular}
\caption{Hyperparameters that were used during training with some typical range of values that we have seen to lead to good performance (see text for a description of each hyperparameter).}
\label{table:hyperparameter_values}
\end{table}

\section{Different connectivities and gatesets}
\label{appendix:connectivity_and_gateset}

\begin{figure}[ht!]
    \centering
    \includegraphics[width=0.44\textwidth]{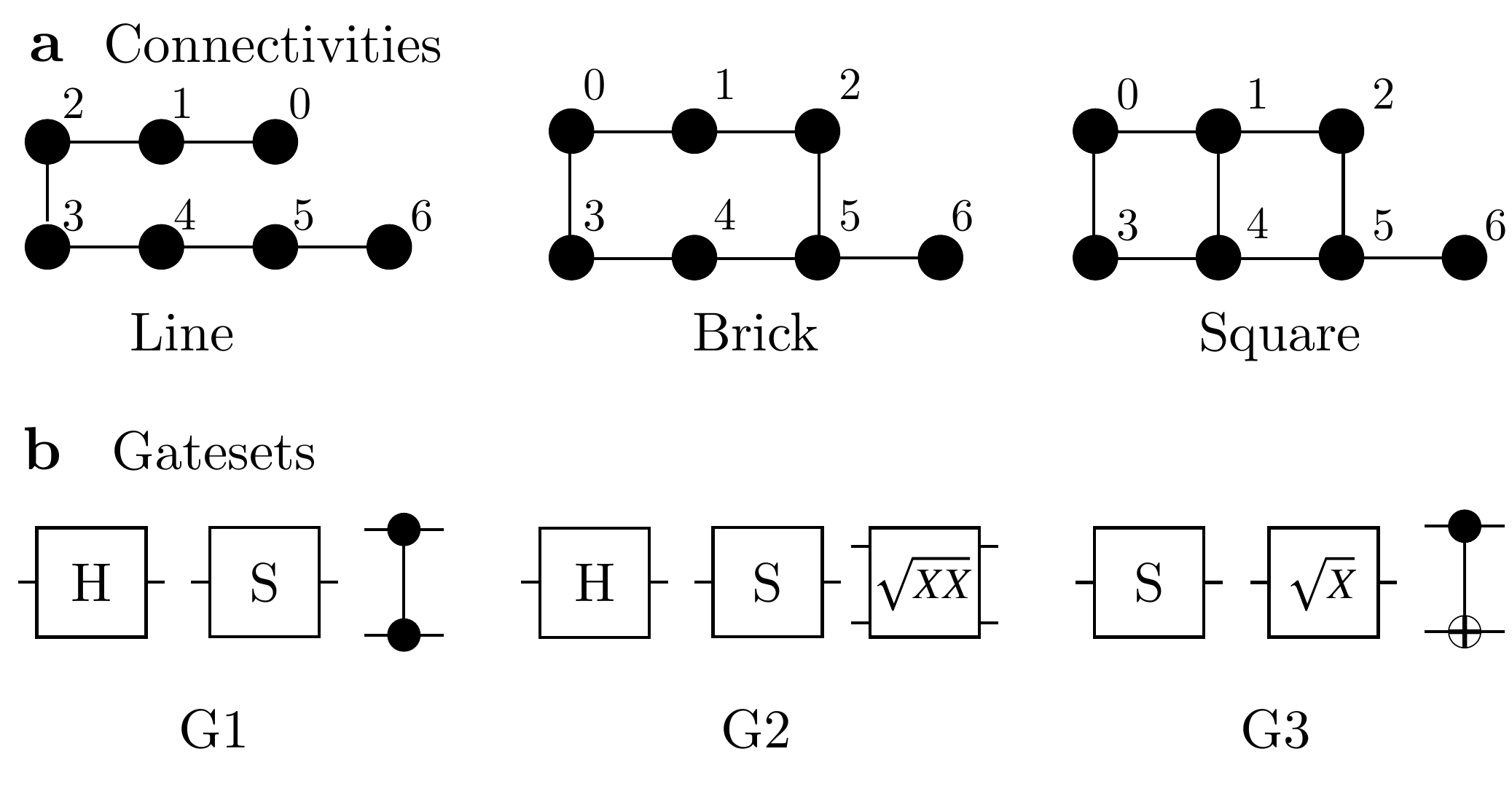}
    \caption{Description of the three different \textbf{a} connectivities and \textbf{b} gatesets that we consider.}
    \label{fig:line_brick_square}
\end{figure}

Here we present results for some other selected gatesets and connectivities to show the flexibility of our approach. For concreteness, we choose codes with parameters $\code{7,1,3}$ and show the shortest encoding circuit for each case. More concretely, we pick three different gatesets and three different connectivities according to Fig.~\ref{fig:line_brick_square}. We have trained 640 agents in every case.

\subsection{Line connectivity:}

\textbf{G1:}

\scalebox{0.9}{
\Qcircuit @C=0.6em @R=0.1em @!R { \\
	 	\nghost{{q}_{0} :  } & \lstick{{q}_{0} :  } & \qw & \ctrl{1} & \gate{\mathrm{H}} & \qw & \control\qw & \qw & \qw & \qw & \qw & \qw & \qw & \qw & \qw & \qw\\
	 	\nghost{{q}_{1} :  } & \lstick{{q}_{1} :  } & \gate{\mathrm{H}} & \control\qw & \control\qw & \gate{\mathrm{H}} & \ctrl{-1} & \qw & \control\qw & \qw & \qw & \ctrl{1} & \gate{\mathrm{H}} & \control\qw & \qw & \qw\\
	 	\nghost{{q}_{2} :  } & \lstick{{q}_{2} :  } & \gate{\mathrm{H}} & \qw & \ctrl{-1} & \ctrl{1} & \gate{\mathrm{H}} & \ctrl{1} & \ctrl{-1} & \ctrl{1} & \gate{\mathrm{H}} & \control\qw & \ctrl{1} & \ctrl{-1} & \qw & \qw\\
	 	\nghost{{q}_{3} :  } & \lstick{{q}_{3} :  } & \gate{\mathrm{H}} & \qw & \qw & \control\qw & \ctrl{1} & \control\qw & \gate{\mathrm{H}} & \control\qw & \qw & \qw & \control\qw & \qw & \qw & \qw\\
	 	\nghost{{q}_{4} :  } & \lstick{{q}_{4} :  } & \gate{\mathrm{H}} & \qw & \qw & \qw & \control\qw & \ctrl{1} & \qw & \qw & \qw & \qw & \qw & \qw & \qw & \qw\\
	 	\nghost{{q}_{5} :  } & \lstick{{q}_{5} :  } & \gate{\mathrm{H}} & \ctrl{1} & \qw & \qw & \qw & \control\qw & \qw & \qw & \qw & \qw & \qw & \qw & \qw & \qw\\
	 	\nghost{{q}_{6} :  } & \lstick{{q}_{6} :  } & \gate{\mathrm{H}} & \control\qw & \qw & \qw & \qw & \qw & \qw & \qw & \qw & \qw & \qw & \qw & \qw & \qw\\
\\ }}

\textbf{G2:}

\scalebox{0.6}{
\Qcircuit @C=0.6em @R=0.1em @!R { \\
	 	\nghost{{q}_{0} :  } & \lstick{{q}_{0} :  } & \multigate{1}{\mathrm{\sqrt{XX}}} & \gate{\mathrm{H}} & \qw & \multigate{1}{\mathrm{\sqrt{XX}}} & \qw & \qw & \qw & \qw & \qw & \qw & \qw\\
	 	\nghost{{q}_{1} :  } & \lstick{{q}_{1} :  } & \ghost{\mathrm{\sqrt{XX}}} & \multigate{1}{\mathrm{\sqrt{XX}}} & \gate{\mathrm{S}} & \ghost{\mathrm{\sqrt{XX}}} & \multigate{1}{\mathrm{\sqrt{XX}}} & \gate{\mathrm{H}} & \qw & \multigate{1}{\mathrm{\sqrt{XX}}} & \qw & \qw & \qw\\
	 	\nghost{{q}_{2} :  } & \lstick{{q}_{2} :  } & \qw & \ghost{\mathrm{\sqrt{XX}}} & \multigate{1}{\mathrm{\sqrt{XX}}} & \gate{\mathrm{H}} & \ghost{\mathrm{\sqrt{XX}}} & \multigate{1}{\mathrm{\sqrt{XX}}} & \gate{\mathrm{S}} & \ghost{\mathrm{\sqrt{XX}}} & \multigate{1}{\mathrm{\sqrt{XX}}} & \qw & \qw\\
	 	\nghost{{q}_{3} :  } & \lstick{{q}_{3} :  } & \multigate{1}{\mathrm{\sqrt{XX}}} & \qw & \ghost{\mathrm{\sqrt{XX}}} & \gate{\mathrm{H}} & \qw & \ghost{\mathrm{\sqrt{XX}}} & \qw & \qw & \ghost{\mathrm{\sqrt{XX}}} & \qw & \qw\\
	 	\nghost{{q}_{4} :  } & \lstick{{q}_{4} :  } & \ghost{\mathrm{\sqrt{XX}}} & \multigate{1}{\mathrm{\sqrt{XX}}} & \qw & \qw & \qw & \qw & \qw & \qw & \qw & \qw & \qw\\
	 	\nghost{{q}_{5} :  } & \lstick{{q}_{5} :  } & \multigate{1}{\mathrm{\sqrt{XX}}} & \ghost{\mathrm{\sqrt{XX}}} & \qw & \qw & \qw & \qw & \qw & \qw & \qw & \qw & \qw\\
	 	\nghost{{q}_{6} :  } & \lstick{{q}_{6} :  } & \ghost{\mathrm{\sqrt{XX}}} & \qw & \qw & \qw & \qw & \qw & \qw & \qw & \qw & \qw & \qw\\
\\ }}

\textbf{G3:}

\scalebox{0.9}{
\Qcircuit @C=0.6em @R=0.1em @!R { \\
	 	\nghost{{q}_{0} :  } & \lstick{{q}_{0} :  } & \ctrl{1} & \qw & \targ & \qw & \qw & \qw & \qw & \targ & \qw & \qw\\
	 	\nghost{{q}_{1} :  } & \lstick{{q}_{1} :  } & \targ & \targ & \ctrl{-1} & \qw & \qw & \targ & \ctrl{1} & \ctrl{-1} & \qw & \qw\\
	 	\nghost{{q}_{2} :  } & \lstick{{q}_{2} :  } & \gate{\mathrm{\sqrt{X}}} & \ctrl{-1} & \ctrl{1} & \gate{\mathrm{\sqrt{X}}} & \targ & \ctrl{-1} & \targ & \qw & \qw & \qw\\
	 	\nghost{{q}_{3} :  } & \lstick{{q}_{3} :  } & \qw & \qw & \targ & \targ & \ctrl{-1} & \qw & \qw & \qw & \qw & \qw\\
	 	\nghost{{q}_{4} :  } & \lstick{{q}_{4} :  } & \gate{\mathrm{\sqrt{X}}} & \ctrl{1} & \qw & \ctrl{-1} & \qw & \qw & \qw & \qw & \qw & \qw\\
	 	\nghost{{q}_{5} :  } & \lstick{{q}_{5} :  } & \qw & \targ & \ctrl{1} & \qw & \qw & \qw & \qw & \qw & \qw & \qw\\
	 	\nghost{{q}_{6} :  } & \lstick{{q}_{6} :  } & \qw & \qw & \targ & \qw & \qw & \qw & \qw & \qw & \qw & \qw\\
\\ }}

\subsection{Brick connectivity:}

\textbf{G1:}

\scalebox{0.9}{
\Qcircuit @C=0.6em @R=0.1em @!R { \\
	 	\nghost{{q}_{0} :  } & \lstick{{q}_{0} :  } & \qw & \control\qw & \qw & \ctrl{1} & \gate{\mathrm{H}} & \control\qw & \qw & \control\qw & \control\qw & \qw & \ctrl{1} & \qw & \qw\\
	 	\nghost{{q}_{1} :  } & \lstick{{q}_{1} :  } & \gate{\mathrm{H}} & \qw & \control\qw & \control\qw & \qw & \qw & \qw & \qw & \ctrl{-1} & \gate{\mathrm{H}} & \control\qw & \qw & \qw\\
	 	\nghost{{q}_{2} :  } & \lstick{{q}_{2} :  } & \gate{\mathrm{H}} & \qw & \ctrl{-1} & \ctrl{3} & \qw & \qw & \qw & \qw & \qw & \qw & \qw & \qw & \qw\\
	 	\nghost{{q}_{3} :  } & \lstick{{q}_{3} :  } & \gate{\mathrm{H}} & \ctrl{-3} & \control\qw & \qw & \qw & \ctrl{-3} & \gate{\mathrm{H}} & \ctrl{-3} & \qw & \qw & \qw & \qw & \qw\\
	 	\nghost{{q}_{4} :  } & \lstick{{q}_{4} :  } & \gate{\mathrm{H}} & \qw & \ctrl{-1} & \qw & \ctrl{1} & \qw & \qw & \qw & \qw & \qw & \qw & \qw & \qw\\
	 	\nghost{{q}_{5} :  } & \lstick{{q}_{5} :  } & \gate{\mathrm{H}} & \ctrl{1} & \qw & \control\qw & \control\qw & \qw & \qw & \qw & \qw & \qw & \qw & \qw & \qw\\
	 	\nghost{{q}_{6} :  } & \lstick{{q}_{6} :  } & \gate{\mathrm{H}} & \control\qw & \qw & \qw & \qw & \qw & \qw & \qw & \qw & \qw & \qw & \qw & \qw\\
\\ }}

\textbf{G2:}

\scalebox{0.7}{
\Qcircuit @C=0.6em @R=0.1em @!R { \\
	 	\nghost{{q}_{0}:} & \lstick{{q}_{0}:} & \multigate{1}{\mathrm{\sqrt{XX}}} & \gate{\mathrm{H}} & \multigate{3}{\mathrm{\sqrt{XX}}} & \gate{\mathrm{H}} & \qw & \multigate{3}{\mathrm{\sqrt{XX}}} & \multigate{1}{\mathrm{\sqrt{XX}}} & \qw & \qw\\
	 	\nghost{{q}_{1}:} & \lstick{{q}_{1}:} & \ghost{\mathrm{\sqrt{XX}}} & \multigate{1}{\mathrm{\sqrt{XX}}} & \ghost{\mathrm{\sqrt{XX}}} & \gate{\mathrm{S}} & \qw & \ghost{\mathrm{\sqrt{XX}}} & \ghost{\mathrm{\sqrt{XX}}} & \qw & \qw\\
	 	\nghost{{q}_{2}:} & \lstick{{q}_{2}:} & \qw & \ghost{\mathrm{\sqrt{XX}}} & \ghost{\mathrm{\sqrt{XX}}} & \multigate{3}{\mathrm{\sqrt{XX}}} & \qw & \ghost{\mathrm{\sqrt{XX}}} & \qw & \qw & \qw\\
	 	\nghost{{q}_{3}:} & \lstick{{q}_{3}:} & \multigate{1}{\mathrm{\sqrt{XX}}} & \qw & \ghost{\mathrm{\sqrt{XX}}} & \ghost{\mathrm{\sqrt{XX}}} & \gate{\mathrm{S}} & \ghost{\mathrm{\sqrt{XX}}} & \qw & \qw & \qw\\
	 	\nghost{{q}_{4}:} & \lstick{{q}_{4}:} & \ghost{\mathrm{\sqrt{XX}}} & \qw & \qw & \ghost{\mathrm{\sqrt{XX}}} & \multigate{1}{\mathrm{\sqrt{XX}}} & \qw & \qw & \qw & \qw\\
	 	\nghost{{q}_{5}:} & \lstick{{q}_{5}:} & \multigate{1}{\mathrm{\sqrt{XX}}} & \qw & \qw & \ghost{\mathrm{\sqrt{XX}}} & \ghost{\mathrm{\sqrt{XX}}} & \qw & \qw & \qw & \qw\\
	 	\nghost{{q}_{6}:} & \lstick{{q}_{6}:} & \ghost{\mathrm{\sqrt{XX}}} & \qw & \qw & \qw & \qw & \qw & \qw & \qw & \qw\\
\\ }}

\textbf{G3:}

\scalebox{0.9}{
\Qcircuit @C=0.6em @R=0.1em @!R { \\
	 	\nghost{{q}_{0} :  } & \lstick{{q}_{0} :  } & \qw & \qw & \targ & \ctrl{3} & \qw & \qw & \qw & \qw\\
	 	\nghost{{q}_{1} :  } & \lstick{{q}_{1} :  } & \qw & \targ & \ctrl{-1} & \qw & \qw & \targ & \qw & \qw\\
	 	\nghost{{q}_{2} :  } & \lstick{{q}_{2} :  } & \gate{\mathrm{\sqrt{X}}} & \ctrl{-1} & \ctrl{3} & \qw & \gate{\mathrm{\sqrt{X}}} & \ctrl{-1} & \qw & \qw\\
	 	\nghost{{q}_{3} :  } & \lstick{{q}_{3} :  } & \qw & \qw & \qw & \targ & \ctrl{1} & \qw & \qw & \qw\\
	 	\nghost{{q}_{4} :  } & \lstick{{q}_{4} :  } & \qw & \qw & \qw & \qw & \targ & \qw & \qw & \qw\\
	 	\nghost{{q}_{5} :  } & \lstick{{q}_{5} :  } & \gate{\mathrm{\sqrt{X}}} & \ctrl{1} & \targ & \qw & \qw & \qw & \qw & \qw\\
	 	\nghost{{q}_{6} :  } & \lstick{{q}_{6} :  } & \qw & \targ & \qw & \qw & \qw & \qw & \qw & \qw\\
\\ }}

\subsection{Square connectivity:}

\textbf{G1:}

\scalebox{0.9}{
\Qcircuit @C=0.6em @R=0.1em @!R { \\
	 	\nghost{{q}_{0} :  } & \lstick{{q}_{0} :  } & \qw & \ctrl{3} & \qw & \qw & \qw & \control\qw & \gate{\mathrm{H}} & \control\qw & \qw & \ctrl{1} & \qw & \qw\\
	 	\nghost{{q}_{1} :  } & \lstick{{q}_{1} :  } & \gate{\mathrm{H}} & \qw & \qw & \ctrl{3} & \control\qw & \ctrl{-1} & \gate{\mathrm{H}} & \qw & \control\qw & \control\qw & \qw & \qw\\
	 	\nghost{{q}_{2} :  } & \lstick{{q}_{2} :  } & \gate{\mathrm{H}} & \qw & \qw & \qw & \ctrl{-1} & \ctrl{3} & \qw & \qw & \qw & \qw & \qw & \qw\\
	 	\nghost{{q}_{3} :  } & \lstick{{q}_{3} :  } & \gate{\mathrm{H}} & \control\qw & \ctrl{1} & \qw & \qw & \qw & \qw & \ctrl{-3} & \qw & \qw & \qw & \qw\\
	 	\nghost{{q}_{4} :  } & \lstick{{q}_{4} :  } & \gate{\mathrm{H}} & \qw & \control\qw & \control\qw & \qw & \qw & \qw & \qw & \ctrl{-3} & \qw & \qw & \qw\\
	 	\nghost{{q}_{5} :  } & \lstick{{q}_{5} :  } & \gate{\mathrm{H}} & \ctrl{1} & \qw & \qw & \qw & \control\qw & \qw & \qw & \qw & \qw & \qw & \qw\\
	 	\nghost{{q}_{6} :  } & \lstick{{q}_{6} :  } & \gate{\mathrm{H}} & \control\qw & \qw & \qw & \qw & \qw & \qw & \qw & \qw & \qw & \qw & \qw\\
\\ }}

\textbf{G2:}

\scalebox{0.7}{
\Qcircuit @C=0.6em @R=0.1em @!R { \\
	 	\nghost{{q}_{0} :  } & \lstick{{q}_{0} :  } & \multigate{1}{\mathrm{\sqrt{XX}}} & \gate{\mathrm{S}} & \multigate{3}{\mathrm{\sqrt{XX}}} & \gate{\mathrm{S}} & \qw & \qw & \multigate{1}{\mathrm{\sqrt{XX}}} & \qw & \qw\\
	 	\nghost{{q}_{1} :  } & \lstick{{q}_{1} :  } & \ghost{\mathrm{\sqrt{XX}}} & \multigate{1}{\mathrm{\sqrt{XX}}} & \ghost{\mathrm{\sqrt{XX}}} & \gate{\mathrm{H}} & \multigate{3}{\mathrm{\sqrt{XX}}} & \gate{\mathrm{H}} & \ghost{\mathrm{\sqrt{XX}}} & \qw & \qw\\
	 	\nghost{{q}_{2} :  } & \lstick{{q}_{2} :  } & \qw & \ghost{\mathrm{\sqrt{XX}}} & \ghost{\mathrm{\sqrt{XX}}} & \multigate{3}{\mathrm{\sqrt{XX}}} & \ghost{\mathrm{\sqrt{XX}}} & \qw & \qw & \qw & \qw\\
	 	\nghost{{q}_{3} :  } & \lstick{{q}_{3} :  } & \multigate{1}{\mathrm{\sqrt{XX}}} & \qw & \ghost{\mathrm{\sqrt{XX}}} & \ghost{\mathrm{\sqrt{XX}}} & \ghost{\mathrm{\sqrt{XX}}} & \qw & \qw & \qw & \qw\\
	 	\nghost{{q}_{4} :  } & \lstick{{q}_{4} :  } & \ghost{\mathrm{\sqrt{XX}}} & \qw & \qw & \ghost{\mathrm{\sqrt{XX}}} & \ghost{\mathrm{\sqrt{XX}}} & \qw & \qw & \qw & \qw\\
	 	\nghost{{q}_{5} :  } & \lstick{{q}_{5} :  } & \multigate{1}{\mathrm{\sqrt{XX}}} & \qw & \qw & \ghost{\mathrm{\sqrt{XX}}} & \qw & \qw & \qw & \qw & \qw\\
	 	\nghost{{q}_{6} :  } & \lstick{{q}_{6} :  } & \ghost{\mathrm{\sqrt{XX}}} & \qw & \qw & \qw & \qw & \qw & \qw & \qw & \qw\\
\\ }}

\textbf{G3:}

\scalebox{0.9}{
\Qcircuit @C=0.6em @R=0.1em @!R { \\
	 	\nghost{{q}_{0} :  } & \lstick{{q}_{0} :  } & \qw & \qw & \ctrl{1} & \gate{\mathrm{\sqrt{X}}} & \ctrl{3} & \qw & \qw & \qw\\
	 	\nghost{{q}_{1} :  } & \lstick{{q}_{1} :  } & \gate{\mathrm{\sqrt{X}}} & \ctrl{3} & \targ & \ctrl{1} & \qw & \qw & \qw & \qw\\
	 	\nghost{{q}_{2} :  } & \lstick{{q}_{2} :  } & \qw & \qw & \qw & \targ & \qw & \qw & \qw & \qw\\
	 	\nghost{{q}_{3} :  } & \lstick{{q}_{3} :  } & \qw & \qw & \qw & \qw & \targ & \ctrl{1} & \qw & \qw\\
	 	\nghost{{q}_{4} :  } & \lstick{{q}_{4} :  } & \qw & \targ & \ctrl{1} & \qw & \qw & \targ & \qw & \qw\\
	 	\nghost{{q}_{5} :  } & \lstick{{q}_{5} :  } & \qw & \qw & \targ & \gate{\mathrm{\sqrt{X}}} & \ctrl{1} & \qw & \qw & \qw\\
	 	\nghost{{q}_{6} :  } & \lstick{{q}_{6} :  } & \qw & \qw & \qw & \qw & \targ & \qw & \qw & \qw\\
\\ }}

\section{Quantum Weight Enumerators of $\code{9,3,3}$}
\label{appendix:QWE_9_3_3}

Here we list the quantum weight enumerators of the 13 families of $\code{9,3,3}$ codes appearing in the main text, ordered accordingly.

\begin{align}
A &=(1, 0, 0, 0, 0, 6, 16, 24, 15, 2)~, \nonumber \\
B &=(1, 0, 0, 40, 162, 480, 952, 1224, 933, 304)~, 
\end{align}

\begin{align}
A &=(1, 0, 0, 0, 2, 4, 12, 28, 17, 0)~, \nonumber \\
B &=(1, 0, 0, 48, 146, 472, 984, 1216, 917, 312)~,
\end{align}

\begin{align}
A &=(1, 0, 0, 0, 2, 2, 16, 28, 13, 2)~, \nonumber \\
B &=(1, 0, 0, 44, 162, 452, 984, 1236, 901, 316)~,
\end{align}

\begin{align}
A &=(1, 0, 0, 0, 1, 3, 18, 26, 12, 3)~, \nonumber \\
B &=(1, 0, 0, 40, 170, 456, 968, 1240, 909, 312)~,
\end{align}

\begin{align}
A &=(1, 0, 0, 1, 0, 3, 16, 27, 15, 1)~, \nonumber \\
B &=(1, 0, 0, 48, 162, 456, 952, 1248, 933, 296)~,
\end{align}

\begin{align}
A &=(1, 0, 0, 0, 0, 4, 20, 24, 11, 4)~, \nonumber \\
B &=(1, 0, 0, 36, 178, 460, 952, 1244, 917, 308)~,
\end{align}

\begin{align}
A &=(1, 0, 0, 0, 1, 5, 14, 26, 16, 1)~, \nonumber \\
B &=(1, 0, 0, 44, 154, 476, 968, 1220, 925, 308)~,
\end{align}

\begin{align}
A &=(1, 0, 0, 1, 0, 1, 20, 27, 11, 3)~, \nonumber \\
B &=(1, 0, 0, 44, 178, 436, 952, 1268, 917, 300)~,
\end{align}

\begin{align}
A &=(1, 0, 1, 0, 0, 4, 13, 28, 17, 0)~, \nonumber \\
B &=(1, 0, 1, 57, 171, 481, 931, 1171, 944, 339)~,
\end{align}

\begin{align}
A &=(1, 0, 0, 0, 4, 0, 12, 32, 15, 0)~, \nonumber \\
B &=(1, 0, 0, 52, 146, 444, 1016, 1228, 885, 324)~,
\end{align}

\begin{align}
A &=(1, 0, 0, 0, 0, 0, 36, 0, 27, 0)~, \nonumber \\
B &=(1, 0, 0, 36, 162, 540, 792, 1404, 837, 324)~,
\end{align}

\begin{align}
A &=(1, 0, 0, 2, 0, 0, 16, 30, 15, 0)~, \nonumber \\
B &=(1, 0, 0, 56, 162, 432, 952, 1272, 933, 288)~,
\end{align}

\begin{align}
A &=(1, 0, 1, 0, 0, 0, 21, 28, 9, 4)~, \nonumber \\
B &=(1, 0, 1, 49, 203, 441, 931, 1211, 912, 347)~,
\end{align}

\section{Distance 5 stabilizer codes}
\label{appendix:distance_5}

Here we show the code families that were found for $d=5$. In order to reduce computational effort, for $n \geq 14$ we ignored \eqref{eq:KL_in_S}, and as a result the codes found in Fig.~\ref{fig:results_families} $n \geq 14$ are only non-degenerate.
Moreover, the increased memory requirements from keeping track of more error operators \eqref{eq:numE} means that the number of agents that can be trained in parallel on a single GPU decreases. For instance, from the 4 agents that we train in parallel, it is rare that any of them finds an encoding sequence that leads to $d=5$ code discovery. In addition, each of these training runs needs 1-4 hours, depending on the code parameters and whether degenerate codes are also targeted.

\begin{figure}[ht!]
    \centering
    \includegraphics[width=0.44\textwidth]{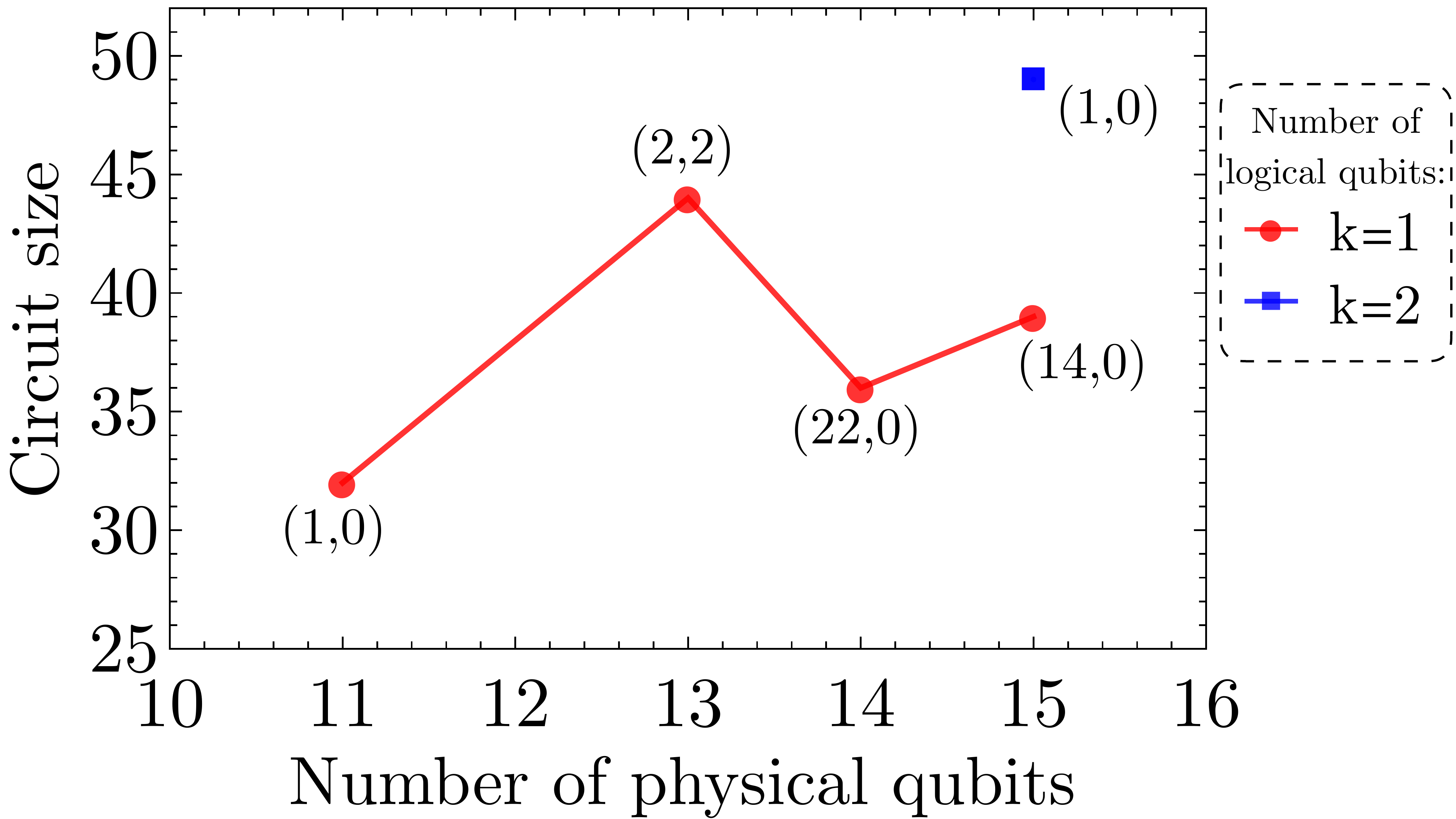}
    \caption{Families of $d=5$ stabilizer codes found with RL. The labels $(x,y)$ indicate the number of non-degenerate $(x)$ and degenerate $(y)$ code families. The circuit size shown is the absolute minimum throughout all families using a directed (CNOT($i<j$)) qubit connectivity.}
    \label{fig:families_d5}
\end{figure}

\section{Encoding circuits}
\label{appendix:encoding_circuits}

Here we show some minimal encoding circuits for selected code parameters. 

$\code{11,1,5}$: 32 gates in a directed all-to-all connectivity:
\scalebox{0.75}{
\Qcircuit @C=0.2em @R=0.1em @!R { \\
	 	\nghost{ } & \ctrl{5} & \gate{\mathrm{H}} & \qw & \qw & \qw & \qw & \ctrl{7} & \qw & \ctrl{8} & \qw & \ctrl{2} & \gate{\mathrm{H}} & \qw & \qw & \ctrl{10} & \qw & \ctrl{9} & \ctrl{1} & \qw & \qw & \qw & \qw & \qw & \qw & \qw & \qw\\
	 	\nghost{ } & \qw & \gate{\mathrm{H}} & \ctrl{9} & \qw & \ctrl{4} & \qw & \qw & \qw & \qw & \ctrl{3} & \qw & \gate{\mathrm{H}} & \ctrl{6} & \ctrl{5} & \qw & \qw & \qw & \targ & \qw & \qw & \qw & \qw & \qw & \qw & \qw & \qw\\
	 	\nghost{ } & \qw & \gate{\mathrm{H}} & \qw & \ctrl{1} & \qw & \ctrl{4} & \qw & \qw & \qw & \qw & \targ & \ctrl{7} & \qw & \qw & \qw & \gate{\mathrm{H}} & \qw & \qw & \ctrl{3} & \qw & \qw & \qw & \qw & \qw & \qw & \qw\\
	 	\nghost{ } & \qw & \qw & \qw & \targ & \qw & \qw & \qw & \ctrl{1} & \qw & \qw & \gate{\mathrm{H}} & \qw & \qw & \qw & \qw & \qw & \qw & \ctrl{7} & \qw & \ctrl{5} & \qw & \ctrl{4} & \qw & \qw & \qw & \qw\\
	 	\nghost{ } & \qw & \qw & \qw & \qw & \qw & \qw & \qw & \targ & \qw & \targ & \qw & \qw & \qw & \qw & \qw & \qw & \qw & \qw & \qw & \qw & \qw & \qw & \qw & \ctrl{1} & \qw & \qw\\
	 	\nghost{ } & \targ & \qw & \qw & \qw & \targ & \qw & \qw & \ctrl{4} & \qw & \qw & \qw & \qw & \qw & \qw & \qw & \ctrl{2} & \qw & \qw & \targ & \qw & \ctrl{1} & \qw & \gate{\mathrm{H}} & \targ & \qw & \qw\\
	 	\nghost{ } & \qw & \qw & \qw & \qw & \qw & \targ & \qw & \qw & \qw & \qw & \qw & \qw & \qw & \targ & \qw & \qw & \qw & \qw & \qw & \qw & \targ & \qw & \qw & \qw & \qw & \qw\\
	 	\nghost{ } & \qw & \qw & \qw & \qw & \qw & \qw & \targ & \qw & \qw & \qw & \qw & \qw & \targ & \qw & \qw & \targ & \qw & \qw & \qw & \qw & \qw & \targ & \qw & \qw & \qw & \qw\\
	 	\nghost{ } & \qw & \qw & \qw & \qw & \qw & \qw & \qw & \qw & \targ & \qw & \qw & \qw & \qw & \qw & \qw & \qw & \qw & \qw & \qw & \targ & \qw & \qw & \qw & \qw & \qw & \qw\\
	 	\nghost{ } & \qw & \qw & \qw & \qw & \qw & \qw & \qw & \targ & \qw & \qw & \qw & \targ & \qw & \qw & \qw & \qw & \targ & \qw & \qw & \qw & \qw & \qw & \qw & \qw & \qw & \qw\\
	 	\nghost{ } & \qw & \qw & \targ & \qw & \qw & \qw & \qw & \qw & \qw & \qw & \qw & \qw & \qw & \qw & \targ & \qw & \qw & \targ & \qw & \qw & \qw & \qw & \qw & \qw & \qw & \qw\\
\\ }}

$\code{15,2,5}$: 49 gates in a directed all-to-all connectivity:

\scalebox{0.6}{
\Qcircuit @C=0.2em @R=0.1em @!R { \\
	 	\nghost{ }  & \ctrl{12} & \gate{\mathrm{H}} & \qw & \qw & \qw & \qw & \qw & \qw & \ctrl{10} & \ctrl{1} & \gate{\mathrm{H}} & \qw & \ctrl{4} & \ctrl{9} & \qw & \qw & \ctrl{5} & \qw & \qw & \qw & \qw & \qw & \qw & \qw & \qw & \qw & \qw & \qw & \qw & \qw\\
	 	\nghost{ }  & \qw & \ctrl{10} & \qw & \qw & \qw & \qw & \qw & \qw & \qw & \targ & \ctrl{5} & \gate{\mathrm{H}} & \qw & \qw & \ctrl{1} & \qw & \qw & \qw & \qw & \qw & \qw & \qw & \qw & \qw & \qw & \qw & \qw & \qw & \qw & \qw\\
	 	\nghost{ }  & \qw & \qw & \gate{\mathrm{H}} & \qw & \qw & \ctrl{5} & \qw & \ctrl{2} & \qw & \ctrl{9} & \qw & \qw & \qw & \qw & \targ & \qw & \qw & \qw & \qw & \qw & \qw & \qw & \ctrl{9} & \qw & \qw & \qw & \qw & \qw & \qw & \qw\\
	 	\nghost{ } &  \qw & \qw & \gate{\mathrm{H}} & \ctrl{6} & \qw & \qw & \ctrl{9} & \qw & \qw & \qw & \qw & \ctrl{3} & \qw & \qw & \gate{\mathrm{H}} & \ctrl{1} & \qw & \ctrl{7} & \qw & \qw & \qw & \qw & \qw & \qw & \qw & \qw & \qw & \qw & \qw & \qw\\
	 	\nghost{ }  & \qw & \qw & \qw & \qw & \qw & \qw & \qw & \targ & \qw & \qw & \qw & \qw & \targ & \qw & \qw & \targ & \qw & \qw & \ctrl{3} & \gate{\mathrm{H}} & \qw & \ctrl{2} & \qw & \qw & \qw & \qw & \qw & \qw & \qw & \qw\\
	 	\nghost{ }  & \qw & \qw & \gate{\mathrm{H}} & \qw & \ctrl{3} & \qw & \qw & \qw & \qw & \qw & \qw & \qw & \qw & \qw & \qw & \qw & \targ & \qw & \qw & \gate{\mathrm{H}} & \ctrl{7} & \qw & \qw & \qw & \qw & \qw & \qw & \qw & \qw & \qw\\
	 	\nghost{ }  & \qw & \qw & \qw & \qw & \qw & \qw & \qw & \qw & \qw & \qw & \targ & \targ & \ctrl{8} & \qw & \gate{\mathrm{H}} & \ctrl{5} & \qw & \qw & \qw & \qw & \qw & \targ & \qw & \ctrl{1} & \qw & \qw & \qw & \qw & \qw & \qw\\
	 	\nghost{ }  & \qw & \qw & \qw & \qw & \qw & \targ & \qw & \gate{\mathrm{H}} & \qw & \qw & \ctrl{5} & \ctrl{1} & \qw & \qw & \ctrl{3} & \qw & \qw & \qw & \targ & \qw & \qw & \qw & \qw & \targ & \ctrl{6} & \ctrl{2} & \qw & \qw & \qw & \qw\\
	 	\nghost{ }  & \qw & \qw & \qw & \qw & \targ & \ctrl{3} & \qw & \qw & \qw & \qw & \qw & \targ & \qw & \qw & \qw & \qw & \qw & \qw & \qw & \qw & \qw & \qw & \qw & \qw & \qw & \qw & \qw & \qw & \qw & \qw\\
	 	\nghost{ }  & \qw & \qw & \qw & \targ & \qw & \qw & \qw & \qw & \qw & \qw & \qw & \qw & \qw & \targ & \qw & \qw & \qw & \qw & \qw & \qw & \qw & \qw & \qw & \qw & \qw & \targ & \ctrl{1} & \ctrl{4} & \qw & \qw\\
	 	\nghost{ }  & \qw & \qw & \qw & \qw & \qw & \qw & \qw & \qw & \targ & \qw & \qw & \qw & \qw & \qw & \targ & \qw & \qw & \targ & \gate{\mathrm{H}} & \qw & \qw & \qw & \qw & \qw & \qw & \qw & \targ & \qw & \qw & \qw\\
	 	\nghost{ }  & \qw & \targ & \qw & \qw & \qw & \targ & \qw & \qw & \ctrl{3} & \targ & \qw & \ctrl{2} & \qw & \qw & \qw & \targ & \gate{\mathrm{H}} & \qw & \qw & \qw & \qw & \qw & \targ & \qw & \qw & \qw & \qw & \qw & \qw & \qw\\
	 	\nghost{}  & \targ & \qw & \qw & \qw & \qw & \qw & \targ & \ctrl{1} & \qw & \qw & \targ & \qw & \qw & \qw & \qw & \qw & \qw & \qw & \qw & \qw & \targ & \qw & \qw & \qw & \qw & \qw & \qw & \qw & \qw & \qw\\
	 	\nghost{}  & \qw & \qw & \qw & \qw & \qw & \qw & \qw & \targ & \qw & \qw & \qw & \targ & \qw & \qw & \qw & \qw & \qw & \qw & \qw & \qw & \qw & \qw & \qw & \qw & \targ & \qw & \qw & \targ & \qw & \qw\\
	 	\nghost{}  & \qw & \qw & \qw & \qw & \qw & \qw & \qw & \qw & \targ & \qw & \qw & \qw & \targ & \qw & \qw & \qw & \qw & \qw & \qw & \qw & \qw & \qw & \qw & \qw & \qw & \qw & \qw & \qw & \qw & \qw\\
\\ }}

$\code{20,13,3}$: 45 gates in an all-to-all connectivity:
\scalebox{0.57}{
\Qcircuit @C=0.2em @R=0.1em @!R { \\
	 	\nghost{ } & \qw & \qw & \qw & \targ & \qw & \ctrl{13} & \qw & \qw & \qw & \qw & \qw & \qw & \qw & \qw & \qw & \qw & \qw & \qw & \qw & \qw & \qw & \qw & \qw & \qw & \qw & \qw & \qw & \qw & \qw & \qw & \qw & \qw & \qw & \qw & \targ & \ctrl{8} & \qw & \qw & \qw\\
	 	\nghost{ } & \qw & \targ & \qw & \qw & \qw & \qw & \qw & \qw & \qw & \qw & \qw & \qw & \qw & \qw & \qw & \qw & \qw & \ctrl{10} & \qw & \qw & \qw & \qw & \qw & \qw & \qw & \qw & \qw & \targ & \qw & \qw & \qw & \qw & \qw & \qw & \qw & \qw & \qw & \qw & \qw\\
	 	\nghost{ } & \qw & \qw & \qw & \qw & \qw & \qw & \qw & \qw & \qw & \qw & \qw & \qw & \qw & \qw & \qw & \targ & \qw & \qw & \qw & \targ & \qw & \qw & \ctrl{13} & \qw & \qw & \qw & \qw & \qw & \qw & \qw & \qw & \qw & \qw & \qw & \qw & \qw & \qw & \qw & \qw\\
	 	\nghost{ } & \qw & \qw & \qw & \qw & \qw & \qw & \qw & \qw & \qw & \qw & \qw & \qw & \qw & \qw & \qw & \qw & \qw & \qw & \qw & \qw & \targ & \qw & \qw & \qw & \qw & \ctrl{16} & \qw & \qw & \qw & \qw & \qw & \qw & \qw & \targ & \qw & \qw & \qw & \qw & \qw\\
	 	\nghost{ } & \qw & \qw & \qw & \qw & \qw & \qw & \qw & \qw & \qw & \qw & \qw & \qw & \qw & \qw & \qw & \qw & \qw & \qw & \targ & \qw & \qw & \ctrl{9} & \qw & \qw & \qw & \qw & \qw & \qw & \qw & \qw & \qw & \qw & \qw & \qw & \qw & \qw & \qw & \qw & \qw\\
	 	\nghost{ } & \qw & \qw & \targ & \qw & \qw & \qw & \qw & \qw & \qw & \qw & \qw & \qw & \qw & \qw & \qw & \qw & \qw & \qw & \qw & \qw & \qw & \qw & \qw & \qw & \qw & \qw & \qw & \qw & \qw & \qw & \targ & \qw & \ctrl{9} & \qw & \qw & \qw & \qw & \qw & \qw\\
	 	\nghost{ } & \qw & \qw & \qw & \qw & \qw & \qw & \qw & \qw & \qw & \qw & \targ & \qw & \qw & \ctrl{9} & \ctrl{1} & \qw & \gate{\mathrm{H}} & \qw & \qw & \qw & \qw & \qw & \qw & \qw & \ctrl{6} & \qw & \qw & \qw & \targ & \qw & \qw & \qw & \qw & \qw & \qw & \qw & \qw & \qw & \qw\\
	 	\nghost{ } & \qw & \qw & \qw & \qw & \qw & \qw & \qw & \qw & \qw & \qw & \qw & \qw & \qw & \qw & \targ & \qw & \qw & \qw & \qw & \qw & \qw & \qw & \qw & \qw & \qw & \qw & \qw & \qw & \qw & \targ & \qw & \qw & \qw & \qw & \qw & \qw & \qw & \qw & \qw\\
	 	\nghost{ } & \qw & \qw & \qw & \qw & \qw & \qw & \targ & \qw & \qw & \ctrl{6} & \qw & \qw & \qw & \qw & \qw & \qw & \qw & \qw & \qw & \qw & \qw & \qw & \qw & \targ & \qw & \qw & \ctrl{8} & \qw & \qw & \qw & \qw & \gate{\mathrm{H}} & \qw & \qw & \qw & \targ & \qw & \qw & \qw\\
	 	\nghost{ } & \qw & \qw & \qw & \qw & \qw & \qw & \qw & \qw & \qw & \qw & \qw & \qw & \qw & \qw & \targ & \qw & \qw & \qw & \qw & \qw & \qw & \qw & \qw & \qw & \qw & \qw & \qw & \qw & \qw & \qw & \qw & \ctrl{7} & \qw & \qw & \qw & \qw & \qw & \qw & \qw\\
	 	\nghost{  } & \qw & \qw & \qw & \qw & \targ & \qw & \qw & \ctrl{8} & \qw & \qw & \qw & \qw & \qw & \qw & \qw & \qw & \ctrl{4} & \qw & \qw & \qw & \qw & \qw & \qw & \qw & \qw & \qw & \qw & \qw & \qw & \qw & \qw & \qw & \qw & \qw & \qw & \qw & \qw & \qw & \qw\\
	 	\nghost{  } & \qw & \qw & \qw & \qw & \qw & \qw & \qw & \qw & \qw & \qw & \qw & \qw & \targ & \qw & \qw & \qw & \qw & \targ & \qw & \qw & \qw & \qw & \qw & \qw & \qw & \qw & \qw & \qw & \qw & \qw & \qw & \qw & \qw & \qw & \qw & \qw & \qw & \qw & \qw\\
	 	\nghost{  } & \qw & \qw & \qw & \qw & \qw & \qw & \qw & \qw & \targ & \qw & \qw & \ctrl{7} & \qw & \qw & \qw & \qw & \qw & \qw & \qw & \qw & \qw & \qw & \qw & \qw & \targ & \qw & \qw & \ctrl{-11} & \qw & \qw & \qw & \qw & \qw & \qw & \qw & \qw & \qw & \qw & \qw\\
	 	\nghost{  } & \qw & \qw & \qw & \qw & \qw & \targ & \qw & \qw & \qw & \qw & \qw & \qw & \qw & \qw & \qw & \qw & \qw & \qw & \qw & \qw & \qw & \targ & \qw & \qw & \qw & \qw & \qw & \qw & \ctrl{-7} & \qw & \qw & \qw & \qw & \qw & \qw & \qw & \qw & \qw & \qw\\
	 	\nghost{  } & \qw & \qw & \qw & \qw & \qw & \qw & \qw & \qw & \qw & \targ & \qw & \qw & \qw & \qw & \qw & \qw & \targ & \gate{\mathrm{H}} & \qw & \qw & \qw & \qw & \qw & \qw & \qw & \qw & \qw & \qw & \qw & \qw & \qw & \qw & \targ & \ctrl{-11} & \ctrl{-14} & \ctrl{2} & \ctrl{4} & \qw & \qw\\
	 	\nghost{  } & \qw & \qw & \qw & \qw & \qw & \qw & \qw & \qw & \qw & \qw & \qw & \qw & \qw & \targ & \qw & \qw & \qw & \qw & \qw & \qw & \qw & \qw & \targ & \qw & \qw & \qw & \qw & \qw & \qw & \qw & \qw & \qw & \qw & \qw & \qw & \qw & \qw & \qw & \qw\\
	 	\nghost{  } & \gate{\mathrm{H}} & \qw & \ctrl{-11} & \qw & \ctrl{-6} & \qw & \qw & \qw & \ctrl{-4} & \qw & \qw & \qw & \ctrl{-5} & \qw & \qw & \ctrl{-14} & \gate{\mathrm{H}} & \qw & \ctrl{-12} & \qw & \ctrl{-13} & \qw & \qw & \qw & \qw & \qw & \targ & \qw & \qw & \qw & \ctrl{-11} & \targ & \qw & \qw & \qw & \targ & \qw & \qw & \qw\\
	 	\nghost{  } & \gate{\mathrm{H}} & \ctrl{-16} & \qw & \ctrl{-17} & \qw & \qw & \ctrl{-9} & \qw & \qw & \qw & \ctrl{-11} & \qw & \qw & \qw & \ctrl{-8} & \gate{\mathrm{H}} & \qw & \qw & \qw & \ctrl{-15} & \qw & \qw & \qw & \ctrl{-9} & \qw & \qw & \qw & \qw & \qw & \qw & \qw & \qw & \qw & \qw & \qw & \qw & \qw & \qw & \qw\\
	 	\nghost{  } & \qw & \qw & \qw & \qw & \qw & \qw & \qw & \targ & \qw & \qw & \qw & \qw & \qw & \qw & \qw & \qw & \qw & \qw & \qw & \qw & \qw & \qw & \qw & \qw & \qw & \qw & \qw & \qw & \qw & \qw & \qw & \qw & \qw & \qw & \qw & \qw & \targ & \qw & \qw\\
	 	\nghost{  } & \qw & \qw & \qw & \qw & \qw & \qw & \qw & \qw & \qw & \qw & \qw & \targ & \qw & \qw & \qw & \qw & \qw & \qw & \qw & \qw & \qw & \qw & \qw & \qw & \qw & \targ & \qw & \qw & \qw & \ctrl{-12} & \qw & \qw & \qw & \qw & \qw & \qw & \qw & \qw & \qw\\
\\ }}

\section{Noise-aware meta-agent vs an ensemble of agents}
\label{appendix:meta_agent_vs_ensemble}

In this work we have shown that a \textit{single} meta-agent trained on different values of the noise bias parameter can find suitable strategies for all values of such parameter. Here, we want to compare the performance of such meta-agent against an ensemble of agents that each have been trained on a \textit{single} value of the noise bias parameter.

We begin by defining the settings of this experiment in order to make a fair comparison. First, there are 16 possible values of the bias parameter $c_Z = \{ 0.5, 0.6, \dots, 1.9, 2\}$. Since each meta-agent has seen instances of all 16 values, we will only allow the single-$c_Z$ agents to be trained on one sixteenth of the total timesteps than the ones used for each meta-agent. In addition, our main results in Section~\ref{sec:results} are based on the best post-selected meta-agents out of 714 training runs. Therefore, we will train $714 \times 16 = 11 424$ single$-c_Z$ agents to make the comparison. All other hyperparameters will be kept fixed.

\begin{figure*}[ht!]
    \centering
    \includegraphics[width=0.97\textwidth]{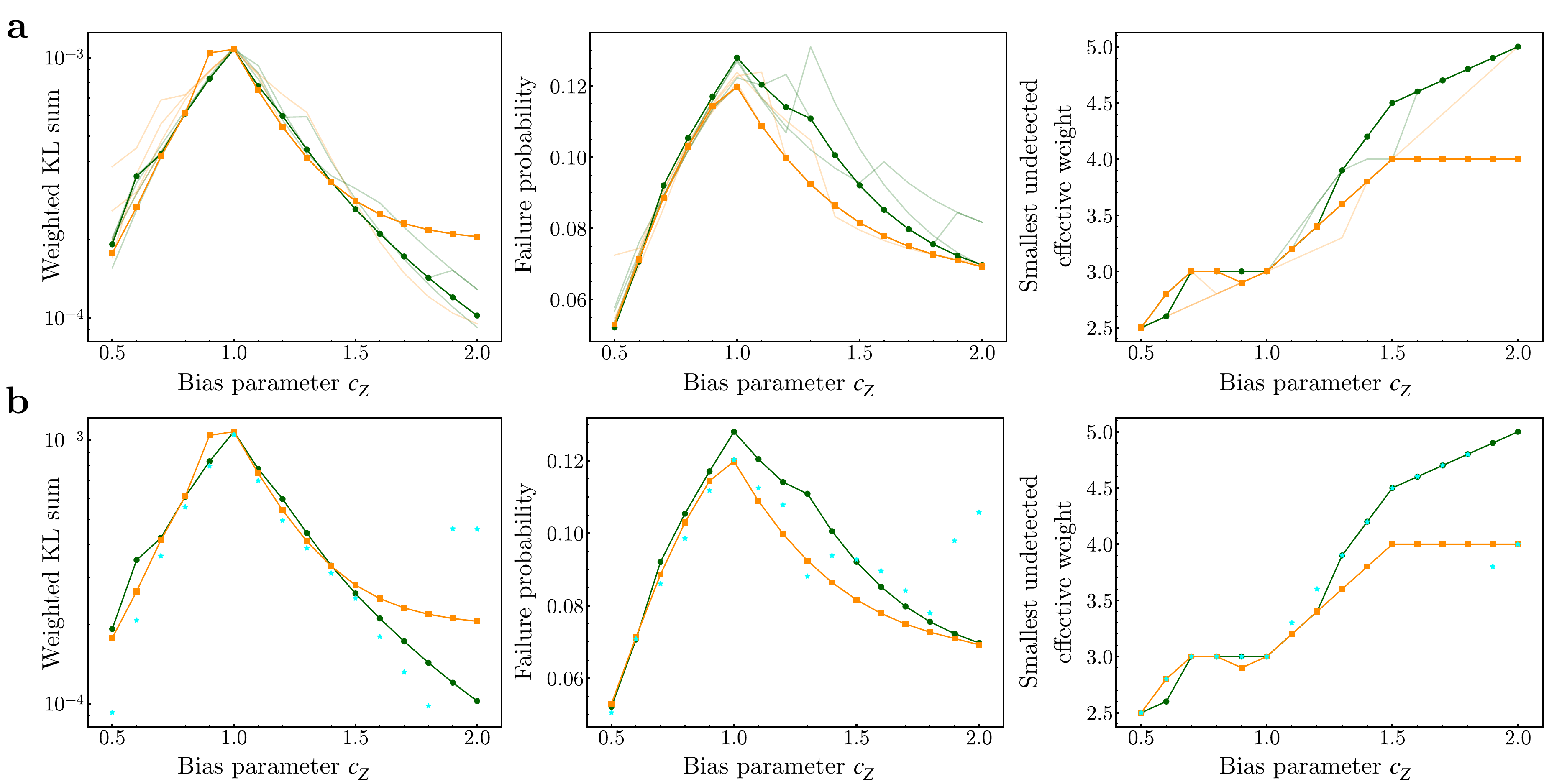}
    \caption{Meta-agent vs ensemble of agents. \textbf{a} The five best performing meta-agents at minimizing the KL sum (green) and at minimizing the failure probability (orange). The faded curves correspond to the second to the fifth best performing meta-agents. \textbf{b} Comparison with the ensemble of agents (cyan star) each trained at a single value of $c_Z$ post-selected to perform best at each metric shown here.}
    \label{fig:meta_agent_vs_agent_ensemble}
\end{figure*}

The results are shown in Fig.~\ref{fig:meta_agent_vs_agent_ensemble}. In panel a we start by recalling the best post-selected meta-agents at minimizing the KL sum (green) and at minimizing the failure probability (orange), also adding the next four best-performing meta-agents for better comparison. In panel b we include with cyan star markers the results for the best single$-c_Z$ agent at every value of $c_Z$. The first stark result is that their performance is rather bad at the extreme values $c_Z=1.9$ and $c_Z=2$ (which some of the meta-agents also struggle with, see panel a). Other than these two, they perform comparably to the best meta-agents, even slightly outperforming them at minimizing the KL sum. However, the meta-agent strategy yields better performance at minimizing the failure probability, which is what is in the end relevant for practical applications. 

\section{Circuit structure of CSS codes}
\label{appendix:CSS}

Here we give a proof of the claim that codes resulting from circuits with an initial block of Hadamard gates on a subset of the qubits and followed by CNOT gates thereafter can only be CSS.

Let us label physical qubits with index $1 \leq q \leq n$ and target a CSS code with parameters $\code{n,k,d}$. Let's assume for simplicity that the initial block of Hadamard gates is applied to qubits $k+1,\dots,k+n_H$, with $n_H < n-k$. The initial tableau of the would-be code reads
\begin{align}
g_1 &= X_{k+1} ~, \nonumber \\
g_2 &= X_{k+2} ~, \nonumber \\
&\cdots \nonumber \\
g_{n_H} &= X_{k + n_H} ~, \nonumber \\
g_{n_H + 1} &= Z_{k + n_H + 1} ~, \nonumber \\
&\cdots \nonumber \\
g_{n - k} &= Z_n ~.
\end{align}
From this moment forward, only CNOT gates are allowed. Let's start by considering what is the effect of a CNOT gate with control qubit inside the H-block, i.e. $\text{control} \in \{ k+1,\dots,k+n_H \}$. For whatever target qubit, what such a CNOT does is populate the target position of the corresponding stabilizer $g_{\text{control}}$ with an X. Subsequent CNOT gates affecting those positions, either as control or target qubits, will either introduce additional X's or simply do nothing. Since $X^2 = 1$, the stabilizers $g_1, g_2, \dots g_{n_H}$ will only ever contain either X's or 1's. Similarly, the effect of CNOTs on stabilizers $g_{n_H + 1}, \dots , g_{n-k}$ is simply populating them with Z's or 1's. Since the set of stabilizer generators can be clearly separated into a subset built with only X's and 1's and another one with only Z's and 1's, such a tableau describes a CSS code.

\section{GPU memory estimation}
\label{appendix:gpu_memory_estimation}

Thanks to the separability of $X$ and $Z$ in the stabilizer generators, the tableaus that we have to simulate are block-diagonal,
\begin{equation}
    \begin{pmatrix}
    g_X & 0 \\
    0 & g_Z \\
    \end{pmatrix}~,
\end{equation}
where $g_X$ is a binary matrix of size $\text{num}(H) \times n$ containing the $X-$type stabilizer generators, and $g_Z$ is of size $(n-k-\text{num}(H)) \times n$ and it contains the representation of the $Z-$type generators. Here, $\text{num}(H)$ is the number of Hadamard gates that are applied at the very beginning (see Fig.~\ref{fig:css_figure}a).

Separability of $X-$ and $Z-$type error detection implies that $g_X$ must detect all $Z-$type errors (by the first KL condition \eqref{eq:KL_anticommute}), and correspondingly for $g_Z$ with $X-$type errors. If the code is degenerate, it must happen that some $X-$type errors are elements of the stabilizer subgroup generated by $g_X$ and likewise for $Z$.

All in all, this means that we can reduce the number of error operators \eqref{eq:numE_CSS} by a factor of 2 (since we use the \textit{same} representation for both $X$ and $Z-$type errors). Each of such error operator is a binary array of size $n$, which amounts to $n$ bytes of memory. 

\end{document}